\documentclass[useAMS,usenatbib,usegraphicx]{mn2e}
\usepackage{journals}
\usepackage{times}
\usepackage{color}
\title[Weighing dark matter in ellipticals]
{Dark matter in
elliptical galaxies: II. Estimating the mass within the virial radius}

\author[Gary A. Mamon
and Ewa L. {\L}okas]{Gary A.
    Mamon$^{1,2}$\thanks{E-mail: gam@iap.fr} and Ewa
    L. {\L}okas$^3$\thanks{E-mail: lokas@camk.edu.pl}\\ $^1$Institut
d'Astrophysique de
    Paris (UMR 7095: CNRS \& Universit\'e Pierre \& Marie Curie), 
98 bis Bd Arago, F--75014 Paris, France \\
$^2$GEPI
    (UMR 8111: CNRS \& Universit\'e Denis Diderot), Observatoire de Paris,
    F--92195 Meudon, France\\
$^3$Copernicus Astronomical Center, Bartycka 18,
    PL--00716 Warsaw, Poland\\
}

\date{Accepted ???. Received ????; in original form ????}
\pubyear{2005}

\begin{document}

\maketitle

\begin{abstract}

Elliptical galaxies are modelled with a a 4-component model: S\'ersic stars,
$\Lambda$CDM dark matter (DM), 
a $\beta$-model 
for the 
hot gas and a central black hole, with the aim of establishing how accurately
can  one measure the total mass within their virial radii. 

DM is negligible in
the inner regions, which are dominated by stars and the central black
hole. This prevents any kinematical estimate (using a Jeans analysis) of the
inner slope of the DM density profile. The gas fraction rises, but the baryon
fraction decreases with radius, at least out to 10 effective radii
($R_e$). Even with line-of-sight velocity dispersion (VD) measurements at 4 or
$5\, R_e$ with $20 \, \rm km \, s^{-1}$ 
accuracy and perfectly known velocity anisotropy, the
total mass within the virial radius ($r_v \equiv r_{200}$) is uncertain by a
factor over 
3. The DM distributions found in $\Lambda$CDM simulations 
appear inconsistent with the low VDs measured by
Romanowsky et al. (2003) of planetary nebulae between 2 and $5\, R_e$, which
typically (but not always) 
imply very low $M/L$s, and a baryon fraction within $r_v$ that is greater
than 
the universal value. Replacing the NFW DM model by the new model of Navarro
et al. (2004) decreases slightly the VD at a given radius. So, given the
observed VD measured at $5\, R_e$, the inferred $M/L$ within $r_v$ is 40\%
larger 
than predicted with the NFW model. Folding in the slight (strong) radial
anisotropy found in $\Lambda$CDM (merger) simulations, which is well modelled
(much 
better than with the Osipkov-Merritt formula) with $\beta(r) = 1/2 \,r/(r+a)$, 
the
inferred $M/L$ within $r_v$ is 1.6 (2.4) times higher than for the
isotropic NFW model. Thus, the DM model and radial anisotropy can partly
explain the low PN VDs, but not in full.
The logarithmic slope of the VD at radii of 1 to $5\,R_e$, which is 
insensitive to
radius,
is another measure of the DM mass
within the virial radius, but it is similarly affected by the \emph{a priori}
unknown DM mass
profile and stellar velocity anisotropy.

Some of the
orbital solutions produced by Romanowsky et al. (2003) indicate that NGC~3379
has a dark matter
content at least as large as cosmologically predicted, and the lower
$M/L$s of 
most of their solutions lead to a baryonic fraction within $r_v$
that is larger than the universal value.
In an appendix, single integral
expressions are derived for the VDs in terms of general radial profiles for
the tracer density and total
mass, for various anisotropic models (general constant anisotropy, 
radial, Osipkov-Merritt, and the
model above).

\end{abstract}

\begin{keywords}
galaxies: elliptical, lenticular, cD --- galaxies:
haloes --- galaxies: structure --- galaxies: kinematics and
dynamics --- methods: analytical
\end{keywords}

\section{Introduction}
\label{intro}

Whereas much work has been devoted to constraining the distribution of dark
matter in spiral galaxies from a multi-component (disk, bulge, halo, and gas)
modelling of their rotation curves (e.g. \citealp{PSS96,SB00}), there have
not been analogous analyses for elliptical galaxies, where the components are
stars, dark matter, hot gas and the central black hole. 
Such an analysis is much more difficult for elliptical galaxies, which 
contrary to spirals, have little rotation \citep{Illingworth77}, so that one
cannot directly infer the mass profiles from circular velocities (assuming
nearly spherical mass distributions).

Instead, one has to analyse the velocity dispersions as a function of
position, and this analysis involves solving the Jeans equation, which in
spherical symmetry, assuming no streaming motions (including rotation), is
\begin{equation}
{{\rm d}(\ell \sigma_r^2) \over {\rm d}r} 
+ 2\,{\beta(r) \over r} \,\ell\sigma_r^2 = -
\ell(r)\,
{GM(r)\over r^2} \ ,
\label{jeans1}
\end{equation}
where $\ell$ is the luminosity density of the galaxy,
$\sigma_r$ its radial velocity dispersion, and
where the anisotropy parameter is 
\begin{equation}
\beta = 1 - {\sigma_t^2 \over\sigma_r^2} \ ,
\label{anisdef}
\end{equation}
with
$\sigma_t = \sigma_\theta = \sigma_\phi$ the 1D tangential velocity dispersion,
so that $\beta = 0$ corresponds to isotropy, $\beta = 1$ is fully radial
anisotropy, and $\beta \to -\infty$ is fully tangential anisotropy.

The luminosity density is easily obtained by deprojecting the surface
brightness profile, and the situation is simplified by 
the recent consensus on the applicability to virtually all elliptical
galaxies \citep*{CCDO93,BCDP02}
of the generalisation (hereafter, S\'ersic law) of the $R^{1/4}$ law
\citep{deVaucouleurs48} proposed
by \cite{Sersic68}, which can be written as
\begin{equation}
I(R) = I_0 \exp \left [-\left ({R \over a_S}\right )^{1/m} \right ] \ ,
\label{Isersic}
\end{equation}
where $I$ is the surface brightness, $a_S$ the S\'ersic scale parameter, and
$m$ the S\'ersic shape parameter, with $m=4$ recovering the $R^{1/4}$ law.
Moreover, strong correlations have been reported between the shape parameter
$m$ and either luminosity or effective (half projected light)
radius $R_e$ (\citealp{CCDO93}; \citealp{PS97}, and references
therein; \citealp{GG03}).

A serious difficulty in modelling the internal kinematics of elliptical
galaxies and clusters of galaxies, considered to be  spherical
non-rotating systems,
is that the Jeans equation~(\ref{jeans1}) has two
unknowns:
the radial profiles of total mass distribution and velocity anisotropy
and this \emph{mass / anisotropy degeneracy} requires some assumption on anisotropy
to recover the total mass distribution.
One solution is to analyse the velocity profiles or at least the 4th order
velocity moments  
(\citealp{Merritt87,RW92,Gerhard93,LM03}; \citealp*{KBM04}).

One can go further in modelling the internal kinematics of elliptical
galaxies or clusters of galaxies with the
\cite{Schwarzschild79} orbit modelling method, which 
supposes a form of the potential and minimises
the differences between the observations (maps of 
surface brightness, mean velocity, and velocity dispersion, or even of
velocity profiles) 
and their predictions obtained by suitable projections of
linear combinations of a set of
orbits that form a basis in energy -- angular momentum phase space.
One can then iterate over the form of the potential to see if one finds
significantly better fits to the observational data.
Similarly, \cite{GJSB98} suggested working with a set of distribution
functions, which one hopes 
forms an adequate basis, and one then minimises 
the linear combination of the distribution
functions that match best the observations.
However, such recent analyses rarely place useful constraints on the
gravitational potential (with no priors), with the exception of
\cite{KSGB00}, who find that some elliptical galaxies are
consistent with constant mass-to-light ratios while others show mass rising
faster than light.

The key issue is to obtain more distant tracers of the gravitational
potential.
Two such studies, by \cite{Mendez+01} and \cite{Romanowsky+03},
of the distribution of the line-of-sight
velocities of planetary nebulae (hereafter, PNe) on the outskirts of
a total of four moderately luminous ($L_*$) nearby
elliptical galaxies indicate fairly rapidly decreasing PN velocity dispersion
profiles.
\citeauthor{Romanowsky+03} carefully analyse one of their galaxies (NGC~3379)
with the Schwarzschild method,
and their favourite conclusion is that the dark matter content
of these galaxies is low at 5 effective radii, and extrapolates to a
very low mass-to-blue-light ratio of $\Upsilon_B = 33$ at 120 kpc, which they
consider as the virial radius.
This low mass-to-light ratio at the virial radius (hereafter, \emph{virial
mass-to-light ratio}) appears to be at strong variance with the
cosmological predictions: 
Two recent works suggest that $\Upsilon_B$ has a non-monotonic
variation with mass (or luminosity), with a minimum value around 80--100 for
luminosities $L_B \approx 10^{9-11}\,{\rm L_\odot}$ \citep{MH02,YMvdB03},
both studies predicting $\Upsilon_B \simeq 100\,h_{70}$ for $L_B = L_{B,*} =
1.88\times 10^{10} L_\odot$ \citep{ML05a}.

There are various alternatives to estimating the mass profiles of elliptical
galaxies through the Jeans equations, in particular modelling the X-ray
emission arising from hot gas assumed in hydrostatic equilibrium in the
gravitational potential, through the effects of gravitational lensing and
through the kinematic analysis of galaxy satellites.

The
analyses of diffuse X-ray emission in elliptical galaxies have the advantage
that the equation of hydrostatic equilibrium, which is the gas equivalent of
the Jeans equation~(\ref{jeans1}),
 has no anisotropy term within
it, so in spherical symmetry, one can easily derive the total mass distribution.
However, the derivation of the total mass profile requires measuring the
temperature profile and its gradient, and 
unfortunately, even with the two new generation X-ray telescopes
{\sf XMM-Newton} and 
{\sf Chandra}, it is difficult to achieve such measurements beyond 
half the virial radius for galaxy clusters \citep{PAP05,Vikhlinin+05}, and even 
much less for elliptical galaxies.
Moreover, the X-ray emission from elliptical galaxies is the combination of
two components: diffuse hot gas swimming in the gravitational potential as well
as direct emission from individual stars, and it is highly difficult to
disentangle the two (see \citealp{BB01}).
Assuming that all the X-ray emission is due to the diffuse hot gas,
\cite{LW99} provide interesting constraints on the dark matter in luminous
($L>L_*$) ellipticals, with a dark matter mass
fraction of $\ga 20\%$ (39\%) at $R=R_e$ ($5\,R_e$).

Weak gravitational lensing is yet another avenue to analyse the gravitational
potential of elliptical galaxies. Since the signal is much too weak for
individual galaxies, one has to stack thousands of galaxies together.
In this manner, \cite{WKLC01} find that the gravitational lensing
shear falls off with angular distance as expected for a structure whose
circular velocity is roughly independent of radius, which suggests
appreciable amounts of dark matter at large radii. Moreover, from their
analysis of SDSS galaxies, \cite{GS02} find that $L_*$ galaxies have
mass-to-light ratio $\Upsilon$ of order of 100 in the $B$ band, again
implying substantial dark matter, since the stellar contribution to the
mass-to-light ratio is thought to be only 
$\Upsilon_* \simeq 8$ in the blue band
(see e.g. \citealp{GKSB01}).
Recently, \cite*{HYG04} derived from the gravitational lensing 
shear profile of RCS galaxies, 
a \citeauthor*{NFW95} (hereafter, NFW) mass within the virial radius,
corresponding to $\Upsilon_B \simeq 59\pm5$.

The constraints from the internal kinematics of galaxy satellites, pioneered
by \cite{ZW94}, are now showing consistency with the $\Lambda$CDM models
\citep{Prada+03}. However, one needs to stack the data from many galaxies, so
errors in stacking can accumulate, and the method is very
sensitive to the correct
removal of interlopers (\citeauthor{Prada+03}).

The constraints on the dark matter are greatly strengthened when combining
the internal kinematics with either the X-ray or gravitational lensing
approach.
Combining internal kinematics with X-rays, \cite{LW99} are able to constrain
the dark matter contents of ellipticals, although they make use of the not
very realistic Osipkov-Merritt \citep{Osipkov79,Merritt85_df} anisotropy (see
Fig.~\ref{betasims} below).
Combining internal kinematics with the constraints from
strong gravitational lensing,
\cite{TK02,TK04} are 
able to constrain the slope of the inner density
profile of the dark matter.

On the theoretical side,
large-scale dissipationless cosmological $N$-body simulations have
recently
reached enough mass and spatial resolution that there appears to be a
convergence on the structure and internal kinematics of the bound
structures,
usually referred to as halos, in the simulations.
In particular, the density profiles have
an outer slope of $\simeq -3$ and an inner slope between $-1$ 
(\citealp*{NFW95,NFW96})
and $-3/2$ \citep{FM97,Moore+99,Ghigna+00}.
In this paper, we consider 
the general formula that \cite{JS00} found to provide
a good fit to simulated halos:
\begin{equation}
\rho(r) \propto \left ({r\over a_d} \right)^{-\alpha} \,\left [1 + \left
({r\over a_d}\right )\right ]^{\alpha-3} \ ,
\label{rhodark}
\end{equation}
where the absolute value of the inner slope $\alpha = 1$
or
3/2,
and `$d$' stands for
\emph{dark}.
These profiles fit well the density profiles of cosmological simulations out
to the \emph{virial} radius $r_v$, wherein the mean density is
$\Delta \approx 200$ times the critical density of the Universe,\footnote{For
the standard $\Lambda$CDM parameters, 
 $\Omega_m=0.3,\Omega_\lambda=0.7$, one finds \citep{ECF96,LH01} $\Delta=102$,
but many cosmologists prefer 
to work with the value of 200, which is close to the value
of 178 originally derived for the Einstein de Sitter Universe ($\Omega_m=1,
\Omega_\lambda=0$).} 
 and are
characterized by their concentration parameter
\begin{equation}
c={r_v \over a_d} \ .
\label{cdef}
\end{equation}

Recently, a number of numerical 
studies have proposed better analytic fits to the radial profiles of
density \citep{DMS04_rho}, density logarithmic slope
\citep{Navarro+04} or circular velocity \citep{SWTS02,Stoehr06}
profiles of simulated halos.
In particular, the formula of 
\citeauthor{Navarro+04} is attractive because it
converges to a finite central density at very small scales and has an
increasing outer slope with a finite mass. 
Moreover, fitting to the logarithmic slope of the density profile is to be
preferred on the grounds that fitting the density or circular velocity
involves fitting a single or double integral of the density slope, thus
possibly missing details smoothed over in the integrals.
We shall therefore focus on the density profile of \citeauthor{Navarro+04}
(hereafter, Nav04) 
in this work.

Two recent studies have placed interesting constraints on the dark matter
distribution within elliptical galaxies.
\cite*{BSD03} compute the aperture velocity dispersions of models with an NFW
dark matter component and a S\'ersic luminosity component. They find that 
the dark matter content must be either low or with a concentration parameter
roughly 3 times lower than fit in the dissipationless cosmological $N$-body
simulations, for otherwise the fundamental plane of elliptical galaxies is
curved more than allowed by observations.
\cite{Napolitano+05} estimate the mass-to-light ratio gradient for a large
number of observed ellipticals. They find that very luminous ellipticals
appear to
have large $M/L$ gradients consistent with important quantities of dark
matter within the virial radius, while 
the lower luminosity $L \la L_*$ ellipticals
appear to have very low gradients, that can be explained either by a strong
lack of dark matter, or else by a dark matter concentration parameter much
lower ($c \simeq 1-6$) than expected from cosmological $N$-body
simulations ($c \simeq 15$).
It remains to be seen how dependent are the conclusions of
\citeauthor{Napolitano+05} on the kinematical modelling of their ellipticals.

Our basic goal is to place constraints on the total mass distribution of
elliptical galaxies.
In a companion paper (\citealp{ML05a}, hereafter, paper I), we show
that the NFW, 
\citeauthor{JS00} (hereafter, JS--1.5), 
and Nav04 density profiles found in dissipationless cosmological
simulations cannot represent the \emph{total} matter distribution in
elliptical galaxies,
because: 1) the local mass-to-light ratios are then far below the generally
admitted values for the stellar component; and 
2) the aperture and slit-averaged
velocity dispersions are much lower than observed for their luminosities  
(i.e. through the \citealp{FJ76}
relation). 
We also show that the highly concentrated NFW models that X-ray
observers have fit to the total mass density profile of ellipticals are an
artifact of fitting to the combination of a S\'ersic stellar component and an
NFW dark matter component, and that the fits cannot be very good.
Finally, we argue that the very low local mass-to-light ratios and aperture
velocity dispersions found imply that
the stellar component fully dominates the internal kinematics of the inner
few effective radii, suggesting that there is little hope in recovering the
inner slope of the dark matter density profile.
All these conclusions suppose that the total mass density profile 
inside an effective radius does not sharply
steepen to a slope of order $-2$
or steeper.

In the present paper, making no direct assumptions on the total mass density
 profile, we focus on the velocity dispersions and mass profiles
 in the outer parts of elliptical galaxies, from 1 to 5 effective radii.
For this, we 
build a detailed 4-component
model of elliptical 
galaxies, with stars, dark matter, hot diffuse gas and a central black hole
 (as we don't know, \emph{a priori}, whether these last two components affect
 the kinematical analysis of dark matter in ellipticals).
It is very possible that the dark matter reacts to the presence of
 the baryon component, which can be treated by the approximation of adiabatic
 contraction, and recent simulations with gas suggest that the resulting dark
 matter density profile scales as $1/r^2$ over a wide range of radii
 (\citealp{GKKN04} using cosmological simulations, and \citealp{Dekel+05}
 using galaxy merging simulations).
Whereas such an ``isothermal'' distribution is a real possibility for the
 dark matter, it remains unclear how feedback processes from the baryons (as
 modelled by \citealp{CJPS05} for the results given by \citeauthor{Dekel+05})
 may
 affect this result.
We therefore  assume here, as in paper~I, 
that the dark matter distribution follows the more robust predictions
 from \emph{dissipationless} $\Lambda$CDM cosmological $N$-body simulations.

The plan of our paper in as follows:
In Sec.~\ref{basic}, we describe our 4-component model, 
in  Sec.~\ref{UpsRomsec}, we
consider the mass-to-light ratios consistent with the orbit modelling of
\cite{Romanowsky+03} on the stellar and PN kinematics in NGC~3379, and in 
Sec.~\ref{velanis}, we estimate the velocity anisotropy of the particles in
dissipationless cosmological $N$-body simulations.
We study the effects of velocity anisotropy in
Sec.~\ref{aniseffects}, compare the importance of the different components in
Sec.~\ref{components}, and ask whether we can weigh the dark matter component in
Sec.~\ref{weighdm}.
We discuss our results in Sect.~\ref{discus}, and
reflect on the very low mass-to-light ratio reported
by \cite{Romanowsky+03} 
for intermediate luminosity elliptical galaxies.
In Appendix~\ref{appJeans}, we
 derive single quadrature expressions for the radial profiles of the
 line of sight velocity dispersion in terms of general tracer density and total
 mass profiles for four simple anisotropy models.

\section{Basic equations}
\label{basic}

We now highlight the 4-component model that we adopt for elliptical galaxies:
stars, diffuse dark matter, hot gas, and a central black hole. We neglect any
contribution from dark baryons (e.g. MACHOs).

\subsection{Distribution of optical light}
\label{dislight}

We begin with the distribution of optical light, referring the reader to
paper~I for details.
The S\'ersic (eq.~[\ref{Isersic}], see detailed properties in \citealp{GD05})
optical surface brightness profile that 
represents the projected stellar distribution, can be deprojected
according to the
approximation first proposed by \cite{PS97}
\begin{eqnarray}
\ell(r) &=& \ell_1\,\widetilde \ell(r/a_S) \label{nuofr}\\
\widetilde \ell(x) &\simeq&
x^{-p}\,\exp \left (-x^{1/m}\right ) \ , \label{nuofx} \\
\ell_1 &=& \left \{{\Gamma(2m)\over \Gamma[(3-p)\,m]}\right\}\,{I_0 \over
2\,a_S} \ , \label{nu1} \\
p &\simeq& 1.0 - 0.6097/m + 0.05463/m^2 \ , \label{pofm}
\end{eqnarray}
where the last equation is from \cite{LGM99}.

The integrated luminosity corresponding to
equations~(\ref{nuofr}), (\ref{nuofx}) and (\ref{nu1}) is then
(\citeauthor{LGM99})
\begin{eqnarray}
L_3 (r) &=& L\, \widetilde L_3(r/a_S) \label{L3ofr} \\
\widetilde L_3(x) &=& {\gamma \left [(3-p)m, x^{1/m}\right ]\over
\Gamma[(3-p)m]} \label{L3tilde} \ ,
\end{eqnarray}
where the total luminosity of the galaxy is 
\begin{equation}
L = 2\pi m\,\Gamma(2m)\,I_0\, a_S^2 = 4\pi m\,
\Gamma\left [(3-p)\,m\right ]\,\ell_1\,a_S^3\ ,
\label{Ltot}
\end{equation}
as obtained by \cite{YC94} from the S\'ersic surface brightness
profile of  
equation~(\ref{Isersic}), and which matches exactly the total
luminosity obtained  by integration of Lima Neto et al.'s
approximate deprojected profile.

It is useful to express radii in terms of the \emph{effective radius}, $R_e$,
which
is the radius of half projected light, where
\begin{eqnarray}
R_e &=& b^m\,a_S \label{Re}\\
b &\simeq& 2\,m - 1/3 + 0.009876/m \ , \label{bofm}
\end{eqnarray}
where the latter relation is from \cite{PS97}.
%
In paper~I, we showed that  $R_e$ and $m$ are fairly well correlated with
total luminosity:
\begin{eqnarray}
\log h_{70}\,R_e^{\rm (fit)} &\!\!\!\!=\!\!\!\!& 
0.34 + 0.54\,\log L_{10} + 0.25\,\left (\log
L_{10}\right)^2 \ , 
\label{Refit}\\
\log m^{\rm (fit)} &\!\!\!\!=\!\!\!\!&
0.43 + 0.26\,\log L_{10} - 0.044\,\left (\log L_{10} \right)^2 \,,
\label{mfit}
\end{eqnarray}
where $L_{10} = h_{70}^2\,L_B / \left ( 10^{10} {\rm L_\odot} \right)$, 
$R_e^{\rm (fit)}$ is measured
in kpc, and
with $H_0 = 100\,h = 70\,h_{70} \, \rm km \,
s^{-1}
\,Mpc^{-1}$.\footnote{Unless noted otherwise, we adopt $h_{70} = 1$.}
Then, equations~(\ref{Re}) and (\ref{bofm}) lead to
\begin{equation}
a_S \simeq \left [b \left(m^{\rm
(fit)}\right)\right]^{-m^{\rm (fit)}} R_e^{\rm (fit)} \ .
\label{aSfit}
\end{equation}
%

\subsection{The central black hole}

High spatial resolution spectroscopic studies of ellipticals have shown that
they almost
always harbour a supermassive black hole of mass 
0.2\% \citep{Faber+97},
0.3\% \citep{Kormendy+97},
or 
0.6\% \citep{Magorrian+98} that of the stellar component.
We thus define the fraction of black hole mass to stellar mass
\begin{equation}
g_{\rm BH} = {M_{\rm BH} \over \Upsilon_*\,L_3(r_v)} \ .
\label{fracBH}
\end{equation}
We could have defined $g_{\rm BH}$ in terms of the total luminosity instead
of that at the virial radius, but the two differ by typically less than
0.1\%, which is much less than the uncertainty on $g_{\rm BH}$,
and the latter scaling generates simpler equations below.
A recent analysis \citep{HR04} favours 
$g_{\rm BH} = 0.15\%\,\left (M_{\rm bulge} / 6 \times 10^{11}
{\rm M_\odot}\right)^{0.12}$, and we will adopt below $g_{\rm BH} = 0.15\%$.
The precise value of $g_{\rm BH}$ has a negligible effect on the constraints
on dark matter in elliptical galaxies.

\subsection{Scalings of global properties}
\label{scalings}

We adopt a fiducial luminosity of $L = L_*$. In paper~I, we derived, from
\cite{Liske+03}, a blue-band luminosity $L_{*,B} = 
1.88\,10^{10}\,h_{70}^{-2} {\rm L_\odot}$,
i.e. an absolute magnitude of $M_B = -20.24$, using (e.g., \citealp{CBC96})
\begin{equation}
M_B^\odot = 5.45 \ .
\label{MBsol}
\end{equation}
Our choice of $L_B$
translates to $m=3.15$ (eq.~[\ref{mfit}]), $R_e=3.2\,\, h_{70}^{-1} \, \rm
kpc$ (eq.~[\ref{Refit}]) and $a_S=11.6\,h_{70}^{-1} \rm
pc$ (eq.~[\ref{aSfit}]). 

In paper~I, we argued that the blue stellar mass-to-light ratios of elliptical
galaxies lay in the rough range from 5 to 8, where the uncertainty is caused
by the uncertain initial mass function and its
lower and upper mass cutoffs, as well as metallicity and stellar evolution
code.
In the present paper, we adopt the mean: $\Upsilon_{*,B} = 6.5$, unless
specified otherwise. 

We also showed in paper~I that the mass-to-light ratio of the Universe
is
\begin{equation}
\overline \Upsilon_B = \Omega_m\,\Upsilon_{{\rm closure},B} =
{3\,\Omega_m H_0^2 \over 8\pi\,G\,j_B} 
= 390\,h_{70} \ ,
\label{mluniv}
\end{equation}
given the luminosity density found by \cite{Liske+03} and \cite{Blanton+03}
in the {\sf 2dFGRS} and {\sf SDSS} surveys, respectively.

We define the \emph{mass-to-light ratio bias}
\begin{equation}
b_\Upsilon = {\Upsilon \over \overline\Upsilon} \ ,
\label{biasUps}
\end{equation}
where $\Upsilon = M_t/L(r_v)$, with $M_t$ the 
total mass within the virial radius.
If the Universe is unbiased, i.e. the mass-to-light ratio within the virial
radius will be
$\Upsilon_B = \overline \Upsilon_B = 390\,h_{70}$.
The internal kinematics of galaxy clusters are consistent
with the universal mass-to-light ratio (e.g. \citealp{LM03} derive
$\Upsilon_B=351$ for the Coma cluster).
Following the predictions of \cite{MH02} and \cite{YMvdB03}, we will adopt a
standard value $\Upsilon_B = \Upsilon_B^{\rm std} = 100\,h_{70}$.

We also define the \emph{baryon fraction bias}
\begin{equation}
b_b = {f_b \over \overline f_b} = {f_b \over \left
(\Omega_b/\Omega_m\right )}  \ ,
\label{biasbarf}
\end{equation}
where $f_b$ is the baryon fraction within the virial radius, while
$\overline f_b = \Omega_b/\Omega_m \simeq 0.14$ is the mean baryon
fraction of 
the Universe. We used here
the big-bang nucleosynthesis measurement 
$\Omega_b=0.02\,h^{-2} = 0.041$ \citep{O'Meara+01}, which is consistent with
the value 
obtained by \cite{Spergel+03} from the {\sf WMAP} CMB experiment, and  
$\Omega_m = 0.3$.

Given the fractions of mass in stars, the central black hole, hot gas and
dark matter, all at the virial radius:
\begin{eqnarray}
f_* &=& {\Upsilon_* \over b_\Upsilon\,\overline\Upsilon} \ ,\label{fstars}\\
f_{\rm BH} &=& g_{\rm BH}\,f_* \ ,
\label{fBH}\\
f_d &=& 1-f_b = 1 - b_b \,\overline f_b \ ,\label{fdark}\\
f_g &=& 1 - \left (f_* + f_{\rm BH} + f_d \right ) \ , \label{fgas}
\end{eqnarray}
where we assumed that the central black hole originates from baryons (this
assumption has a negligible effect on what follows),
it is easy to show that these two biases are related through
\begin{equation}
b_b\,b_\Upsilon = {1\!+g\!+\!g_{\rm BH} \over \overline f_b}
\,\left ({\Upsilon_* \over
\overline\Upsilon}\right) 
\simeq {1\!+\!g\over \overline f_b}\,
\left ({\Upsilon_* \over
\overline\Upsilon}\right) \ ,
\label{bias2}
\end{equation}
where $g$ is the gas to star ratio within the virial radius, and $\Upsilon_*$
is the mass-to-light ratio of the stellar population, assumed independent of
radius (in conformance with the very weak colour gradients in ellipticals,
e.g. \citealp{Goudfrooij+94}).

If mass is biased  relative to luminosity in elliptical galaxies ($b_\Upsilon
\neq 1$),
we can assume that the baryon fraction is unaffected ($b_b = 1$, and
$g$ must vary according to
eq.~[\ref{bias2}]).
Indeed, it is difficult to conceive of a mechanism that will segregate
baryons from dark matter within a radius as large as the virial radius.
For the gas to star ratio to be positive, one then requires
\begin{equation}
\Upsilon_B > {\Upsilon_{*,B} \over \overline f_b} = 
48\,\left ({\Upsilon_{*,B}\over 6.5}\right ) \ .
\label{gpos}
\end{equation}
In other words, considering more mass in stars requires a larger minimum
total mass at the virial radius for
the baryon fraction to retain its universal value.
Turning the argument around, a low total mass-to-light ratio at the virial
radius would imply that the baryonic fraction is greater than the universal
value, i.e. $b_b > 1$.
Indeed,
the general
equation~(\ref{bias2}) implies
\begin{equation}
b_b > {\Upsilon_* / \Upsilon \over \overline f_b} = 1.0\ {\Upsilon_*/6.5\over
\Upsilon/48} \ ,
\label{bbmin}
\end{equation}
where the equality is for a negligible gas-to-star ratio $g$.


\subsection{Distribution of dark mass}

We consider here three dark matter models: the NFW model with inner slope $-1$,
the generalized NFW model introduced by \cite{JS00} with inner slope $-3/2$
(JS--1.5), and the convergent model of \citeauthor{Navarro+04}
(\citeyear{Navarro+04}, Nav04).
The dark matter density profile can generally be written (see paper I,
especially for the Nav04 model):
\begin{eqnarray}
\rho_d(r) &\!\!\!\!=\!\!\!\!& {c^3\over g_d(c)}\,
\left ({M_d\over 4\pi\,r_v^3}\right)\,
\widetilde \rho_d(r/a_d) \
,
\label{rhodark2}
\\
\widetilde \rho_d(y) &\!\!\!\!=\!\!\!\!& 
\left \{
\!\!
\begin{array}{lll}
y^{-\alpha}\,(1+y)^{\alpha-3} 
& \!\!\!\!\alpha=1\ \hbox{(NFW)}, \ \alpha=3/2\ \hbox{(JS--1.5)} \ , \\
& \\
\exp\left (- {2\mu}\,y^{1/\mu} \right )
& \hbox{(Nav04)} \ ,
\end{array}
\right.
\label{rhotildedark}
\\
g_d(y) &\!\!\!\!=\!\!\!\!& 
\left \{
\!\!\!\!
\begin{array}{ll}
\ln(y+1)-y/(y+1) & \!\!\!\!\hbox{(NFW)} \ ,
\label{gdark}\\
& \\
2\,\left [
\sinh^{-1} \sqrt{y}
-\sqrt{y/(y\!+\!1)} \right ]&
\!\!\!\!\hbox{(JS--1.5)} \ , \\
& \\
{1\over2}\,\left (2\,\mu \right)^{1-3\,\mu}\,
\gamma \left [{3\,\mu},{2\,\mu}\,y^{1/\mu} \right ]
& \!\!\!\!\hbox{(Nav04)} \ ,
\end{array}
\right.
\label{gofc}
\end{eqnarray}
where $\alpha$ is the absolute value of the logarithmic slope of the
inner density profile (for the NFW and JS--1.5 models), $c$ is the
concentration parameter (eq.~[\ref{cdef}]), 
$a_d$ is the radius where the logarithmic slope is equal to $-2$ (NFW,
Nav04) or $-9/4$ (JS--1.5, for which $a_d/2$ is the radius where the slope is
$-2$), $\mu \simeq 6$ (paper~I), and $\sinh^{-1} x =
\ln(x+\sqrt{x^2+1})$ for $x>0$.
%
%
In equation~(\ref{rhodark2}),
$M_d$ is the dark mass within the virial radius, defined
such that the mean \emph{total} density within it
is $\Delta=200$ times the critical density of the Universe, $\rho_{\rm crit} =
3\,H_0^2/(8\,\pi G)$, yielding a virial radius (see \citealp{NFW97})
\begin{eqnarray}
r_v &=& \left ({2\,G\,M_v\over \Delta\,H_0^2}\right )^{1/3} \ ,\nonumber \\
&=& 163 \, h^{-1} \, {\rm kpc}
\,\left ({h\,M_t \over  10^{12}\,{\rm M_\odot}} \right )^{1/3} \ , \nonumber \\
&=& 206\,h_{70}^{-1} \, {\rm kpc}
\,\left ({h_{70} M_t \over  10^{12}\,{\rm M_\odot}} \right )^{1/3} \ ,
\label{rvir}
\end{eqnarray}
for $\Delta = 200$.
\cite{JS00} measured 
the
concentration parameter $c$ (eq.~[\ref{cdef}]) from their
$\Lambda$CDM (with cosmological density parameter 
$\Omega_m=0.3$ and dimensionless cosmological constant
$\Omega_\Lambda=0.7$) simulations, which can be fit by the relations
%
\begin{equation}
c \simeq \left\{
\begin{array}{ll}
\displaystyle
10.2 \,M_{12}^{-0.08}
& (\hbox{NFW}) \ , \\
& \\
\displaystyle
4.9 \,M_{12}^{-0.13}
& (\hbox{JS--1.5}) \ ,
\label{cJS}
\end{array}
\right.
\end{equation}
where
$M_{12} = {h M_t / 10^{12} {\rm M_\odot}}$.
In paper~I, we derived the concentration
parameter of the Nav04 model:
\begin{equation}
c = 8.1\,M_{12}^{-0.11-0.015\,\log M_{12}} \,
\label{cNav04fit}
\end{equation}
which is
similar to the concentration parameters in equations~(\ref{cJS}).
For $L_B=L_{*,B}$, equations~(\ref{Refit}) and (\ref{rvir}) lead to
\begin{equation}
{r_v \over R_e} = 79 \,\left ({\Upsilon_B\over 100}\right)^{1/3} \ .
\label{rviroverRe}
\end{equation}

The cumulative mass of the dark models used here can all be written
\begin{eqnarray}
M_d(r) &=& M_d\,\widetilde M_d(r/a_d) \label{Mdarkofr} \ ,\\
\widetilde M_d(y) &=& {g_d(y) \over g_d(c) } \ ,
\label{mtildedark}
\end{eqnarray}
where $g_d(y)$ is given in equation~(\ref{gdark}).\footnote{Note 
that the definition of $g_d$ is the inverse of the definition
of $g$ given by \cite{LM01} for the NFW model.}



A complication arises because the luminosity at the virial radius is very
close to but not exactly equal to the total luminosity.
We write out the total mass at the
virial radius, both as a mean density threshold, and as the luminosity at the
virial radius times the mass-to-light ratio:
\begin{equation}
M_t = {4\pi\over 3} \,r_v^3\,\Delta\,\rho_{\rm crit} = {\Delta H_0^2
r_v^3\over 2\,G} =
\Upsilon\,L\,\widetilde L_3 (r_v/a_s) \ ,
\label{mtotrvir}
\end{equation}
(where we used eq.~[\ref{biasUps}])
and solve the last equality of
equation~(\ref{mtotrvir}) for $r_v$ (the correction on $r_v$ turns out to be
negligible as the luminosity has nearly fully converged at the virial
radius).

\subsection{Distribution of gas mass}

Elliptical galaxies also shine X-rays.
Both discrete sources and a hot diffuse interstellar medium contribute to
their X-ray emission, and the latter component should contribute
non-negligibly to the mass budget of ellipticals.
Indeed, if elliptical galaxies
are unbiased matter tracers ($b_\Upsilon=1$), then
according to equation~(\ref{fstars}),
stars contribute at the virial radius to a fraction of the total mass of
$f_* = \Upsilon_*/\overline \Upsilon =
0.017\,(\Upsilon_*/6.5)$.
This
is much less than the baryon fraction at the virial radius assuming no baryon
bias ($b_b=1$), i.e.
$f_b = \overline f_b =  0.14\,h_{70}^{-2}$ (Sect.~\ref{scalings}).
So, if ellipticals have the same baryon fraction within their virial radius
as the full
Universe, then the hot gas in ellipticals accounts for $0.14\,h_{70}^{-2}
-0.025\,h_{70}^{-1} =
0.11$ (for $h_{70} = 1$)
of the total mass within the virial radius (assuming no dark baryons), which
is $0.11/0.025 \simeq 4$ times more than stars.
If elliptical galaxies have virial mass-to-light ratios as low as
$\Upsilon_B=100$, then 
the gas-to-star ratio is lower (for fixed baryonic fraction).
According to equation~(\ref{bias2}), for  no baryon
bias ($b_b=1$), the gas dominates the stars
at the 
virial radius ($g>1$) for $\Upsilon_B > 2\Upsilon_{*,B}/\overline f_b = 93
\left(\Upsilon_{*,B}/8\right)$.

\cite{BB01} have fit the X-ray surface brightness profiles of luminous
ellipticals (jointly with a component for discrete sources following the
$R^{1/4}$ law), and find that the usual so called $\beta$-model
\begin{eqnarray}
\rho_g (r) &=& \rho_0 \,\widetilde\rho_g (r/r_c)
\label{rhogas} \\
\widetilde\rho_g (z) &=& \left (1+z^2 \right)^{-3\beta_g/2} \ ,
\label{rhogastilde}
\end{eqnarray}
provides a good representation of the distribution of the hot gas,\footnote{Our
parameters $z$ and $\beta_g$ have
nothing to do with a redshift or velocity
anisotropy, respectively!} with
$\langle \beta_g \rangle = 0.49$ and 
\begin{equation}
r_c \simeq {R_e\over q} \ ,
\label{rcoverRe}
\end{equation}
where $q \simeq 10$.
Using a larger sample of elliptical galaxies, 
\cite{OPC03} find a mean $\beta_g$ of 0.55,
but do not estimate $r_c/R_e$.
These parameters should be considered tentative, as the subtraction of the
hard stellar component is quite uncertain.

As for the other non-stellar components, 
the gas enters our analysis only through its
cumulative mass (see the Jeans equation~[\ref{jeans1}]).
Adopting 
\begin{equation}
\beta_g = 1/2 \ ,
\label{betagfit}
\end{equation}
the cumulative mass distribution arising from equation~(\ref{rhogas}) is,
\begin{eqnarray}
M_g(r) &\!\!\!\!=\!\!\!\!& 
{4\,\pi}\,\rho_0\,r_c^3\,\widetilde M_g
(r/r_c)  
\label{mgas}\\
\widetilde M_g (z) &\!\!\!\!=\!\!\!\!&
{2\over3}\,z\,\left [(1+z^2)^{1/4}-
{}_2F_1 \left({1\over2},{3\over4};{3\over2};-z^2\right)\right] \ ,
\end{eqnarray}
where ${}_2F_1(a,b;c;x) = \Gamma(c)/[\Gamma(b)\Gamma(1\!-\!b)] \int_0^1
t^{b\!-\!1}(1\!-\!t)^{c\!-\!b\!-\!1}(1\!-\!t x)^{-a}$ is a the Hypergeometric
function. 
Note that $\widetilde M_g(1) \simeq 0.239$.

To an accuracy of 2.7\% for all $z$, one has
\begin{equation}
\widetilde M_g (z) \simeq \left [
\left ({1\over 3}\,z^3 \right)^{-\gamma} + 
\left ({2\over3}\,z^{3/2}\right)^{-\gamma}
\right]^{-1/\gamma}
\ ,
\label{mtildegasapx}
\end{equation}
with $\gamma = 2^{1/8} \simeq 1.0905$.

We write the gas mass profile as
\begin{equation}
M_g(r) = M_g(r_v)\,
{\widetilde M_g(r/r_c) \over 
\widetilde M_g(r_v/r_c)}
\label{mgas2}
\end{equation}
and normalise the gas component with the
baryon fraction within the virial radius:
\begin{equation}
{M_g (r_v)
\over
M_t} = f_b-\left(1+g_{\rm BH}\right)\,f_*
=
b_b\,\overline f_b  - \left(1+g_{\rm BH}\right )\,{\Upsilon_* \over \Upsilon}
\ .
\label{gastotot}
\end{equation}

The divergence of the gas mass profile at large radii is very severe ($M_g
\propto r^{3/2}$), leading to a divergent gravitational potential, and to a
local gas fraction much 
too large.
We therefore assume that, beyond the virial radius,
the ratio of local baryon (gas + stellar) mass density  
to local total matter density is equal to
the universal baryon ratio $\overline f_b = \Omega_b/\Omega_m$, yielding
\begin{equation}
\rho_g(r) = {\overline f_b \over 1 - \overline f_b}\,\rho_d(r) -
\Upsilon_*\,\ell(r) \qquad \hbox { for } r>r_v\ ,
\label{rhogasout}
\end{equation}
where $\rho_d$ is the dark matter density.

The local gas density profile will therefore be discontinuous at the virial
radius. Physically, one expects a shock to occur at the interface between the
gas infalling into the galaxy and the gas in equilibrium within the galaxy,
and this shock should occur very close to the virial radius.
With equations~(\ref{rhogas}), (\ref{rhogastilde}),
(\ref{rcoverRe}), (\ref{betagfit}), (\ref{mgas}), (\ref{mtildegasapx}), and
(\ref{gastotot}), the
local gas density profile just within the virial radius is (dropping the
negligible $1+g_{\rm BH}$ term in eq.~[\ref{gastotot}])
\begin{equation}
\rho_g(r_v^-) = 
\left (b_b\,\overline f_b - {\Upsilon_* \over \Upsilon} \right )
{\left (1 + r_v^2/r_c^2 \right)^{-3/4}\over \widetilde M_g (r_v/r_c)}
\,\left ({M_t\over 4\pi r_c^3} \right )\,,
\label{rhorvirminus}
\end{equation}
while with equations~(\ref{rhodark2}), (\ref{rhotildedark}), 
(\ref{gofc}), and (\ref{rhogasout}),
the local gas density just outside the virial radius is
\begin{eqnarray}
\rho_g(r_v^+) &\!\!\!\!=\!\!\!\!& 
\left \{
{\left[c/(c+1)\right]^{3-\alpha}\over g_d(c)}\,
\overline f_b
\left (
{1 -b_b\,
\overline f_b
\over
1 -
\overline f_b}
\right)\, \right.
\nonumber \\
&\!\!\!\!\mbox{}\!\!\!\!& 
\left. - {(r_v/a_S)^3\,\widetilde \ell(r_v/a_S) \over
m\,\Gamma[(3-p)m]\,\widetilde L_3(r_v/a_S)}\,
{\Upsilon_*\over \Upsilon}\right\}\,
\left({M_t\over 4\pi r_c^3} \right )\,.
\label{rhorvirplus}
\end{eqnarray}

According to equations~(\ref{rhorvirminus}) and (\ref{rhorvirplus}), the
discontinuity of the gas density at the virial radius can be written
\begin{eqnarray}
{\rho_g(r_v^+)
\over
\rho_g(r_v^-)} &=&
{1\over b_b \Upsilon\, \overline f_b - \Upsilon_*}\,
{\left (1+z_v^2 \right)^{3/4}\,\widetilde M_g(z_v) \over z_v^3}
\nonumber \\
&\mbox{}& 
\times \left \{
\Upsilon \,\overline f_b\,
\left (
{1 -b_b\,\overline f_b
\over
1 -\overline f_b}
\right)\,
\,{\left [c/(c+1)\right]^{3-\alpha}\over g_d(c)} \right. \nonumber \\
&\mbox{}& 
\qquad - \left.
{\Upsilon_*\over m \Gamma[(3-p)m]}\,
{x_v^3\,\widetilde \ell(x_v) \over \widetilde L_3(x_v)}
\right\} \ ,
\label{jump}
\end{eqnarray}
where
$x_v = {r_v/ a_S}$ and $z_v = r_v/r_c$.
For a wide variety of plausible parameters $\Upsilon_*$, $b_b$, $\Upsilon$,
$\alpha$, $c$, $x_v$ and $z_v$, equation~(\ref{jump}) yields 
density ratios $\rho_g(r_v^-)/\rho_g(r_v^+)$ between 2 and 3.5, in accordance
with 
the standard Rankine-Hugoniot conditions (density ratio 
smaller than 4).

The cumulative gas mass beyond the virial radius is then
\begin{eqnarray}
M_g(r) &=& M_g(r_v) + 
{\overline f_b \over 1 - \overline f_b}\,\left [M_d(r) - M_d(r_v) \right ]
\nonumber \\
&\mbox{}& \qquad\ \ \ \ \, - \Upsilon_* \,\left [L(r) - L(r_v) \right ] \ ,
\label{mgasbigr}
\end{eqnarray}
yielding a normalized gas mass 
satisfying
\begin{eqnarray}
{\widetilde M_g (z) \over
\widetilde M_g (z_v)} - 1
&=& {M_g(r)  \over M_g(r_v)} -1
\nonumber 
\\
&=& 
{f_d \over f_g}\,{\overline f_b \over \left (1 \!-\! \overline f_b\right )}\,
\left [\widetilde M_d \left ({r \over a_d} \right ) - 1 \right ] \nonumber \\
&\mbox{}& \quad 
- {f_*\over f_g}\,\left [{\widetilde L_3 \left (r/a_S \right) \over
\widetilde L_3 \left (r_v / a_S \right) } - 1 \right ]
\ ,
\label{mtildegas2}
\end{eqnarray}
where the second term is negligible (as the integrated luminosity has almost
fully converged at the virial radius).

\section{Other astronomical inputs}
\subsection{Lower bound on the virial mass-to-light ratio}
\label{UpsRomsec}

Our lower bound on the mass-to-light ratio within the virial radius is taken
from the work of \cite{Romanowsky+03}, who performed, for the nearby giant
elliptical, NGC~3379, an orbit modelling of
the velocity field of the planetary nebulae combined with the radial profiles
of the stellar surface
brightness and line-of-sight velocity  dispersion.
Although they quote a mass-to-light ratio at 120 kpc, which they estimate as
the virial radius, of $\Upsilon_B
= 33\pm14$, we find
\begin{equation}
\Delta(r) = 2\,{\Upsilon_B\,G\,L_B\over H_0^2\,r^3} = 507 \ ,
\end{equation}
for $r = 120\,\rm kpc$, $\Upsilon_B=33$, and $L_B = 1.51\times
10^{10}\,L_\odot$ (from \citeauthor{Romanowsky+03}'s $M_B = -20.0$ and
eq.~[\ref{MBsol}]). 
So clearly the radius of overdensity 200 is much larger than 120 kpc, and
since dark matter is believed to be more extended than luminous matter (e.g.,
Fig.~\ref{figcomponents}), the mass-to-light ratio should be larger than
33.


We extrapolate \citeauthor{Romanowsky+03}'s mass-to-light ratio out to the
virial radius, as follows.
In their orbit modelling of
NGC~3379, 
\citeauthor{Romanowsky+03} (supporting online material)  
write the density profile as the sum of a luminous
\cite{Hernquist90} component and a dark NFW component:
\begin{equation}
\rho(r) = {a^2\,v_*^2 \over 2\,\pi\,G\,r\,(r+a)^3} + 
{r_s\,v_s^2\over 4\,\pi\,G\,r\,(r+r_s)^2} \ .
\end{equation}
Writing $r = a\,x$ and $r_s = a\,x_s$, it is easy to show that
the mass enclosed within radius $r$ satisfies
\begin{equation}
{G M(r)\over a v_*^2} =
\left \{\! \left (\!{x\over x\!+\!1}\right)^2\! + 
\left ({v_s\over v_*}\right)^2 \! x_s\left [\ln \left(1\!+\!{x\over
    x_s}\right) -{x\over x\!+\!x_s} \right ] \right\}.
\label{MRom03ofr}
\end{equation}

Expressing the mean density at the virial radius as $\Delta$ times the
critical density 
of the Universe, one needs to solve
for 
$x_v = r_v/a = c\,x_s$ the equation 
\begin{eqnarray} 
&&\left \{\left  ({x_v \over x_v+1}\right)^2 + 2\,
\left ({v_s\over v_*}\right)^2\,x_s\,\left [\ln \left(1+{x_v\over
    x_s}\right) -{x_v\over x_v+x_s} \right ] \right\} \nonumber \\
&\mbox{}& \quad 
= \left ({\Delta\over 2}\right )\,\left ({H_0\,a\over v_* }\right)^2 \ ,
\label{xveq}
\end{eqnarray}
where $r_{\rm vir}$ is the virial radius for the total matter.

\begin{figure}
\resizebox{\hsize}{!}{\includegraphics{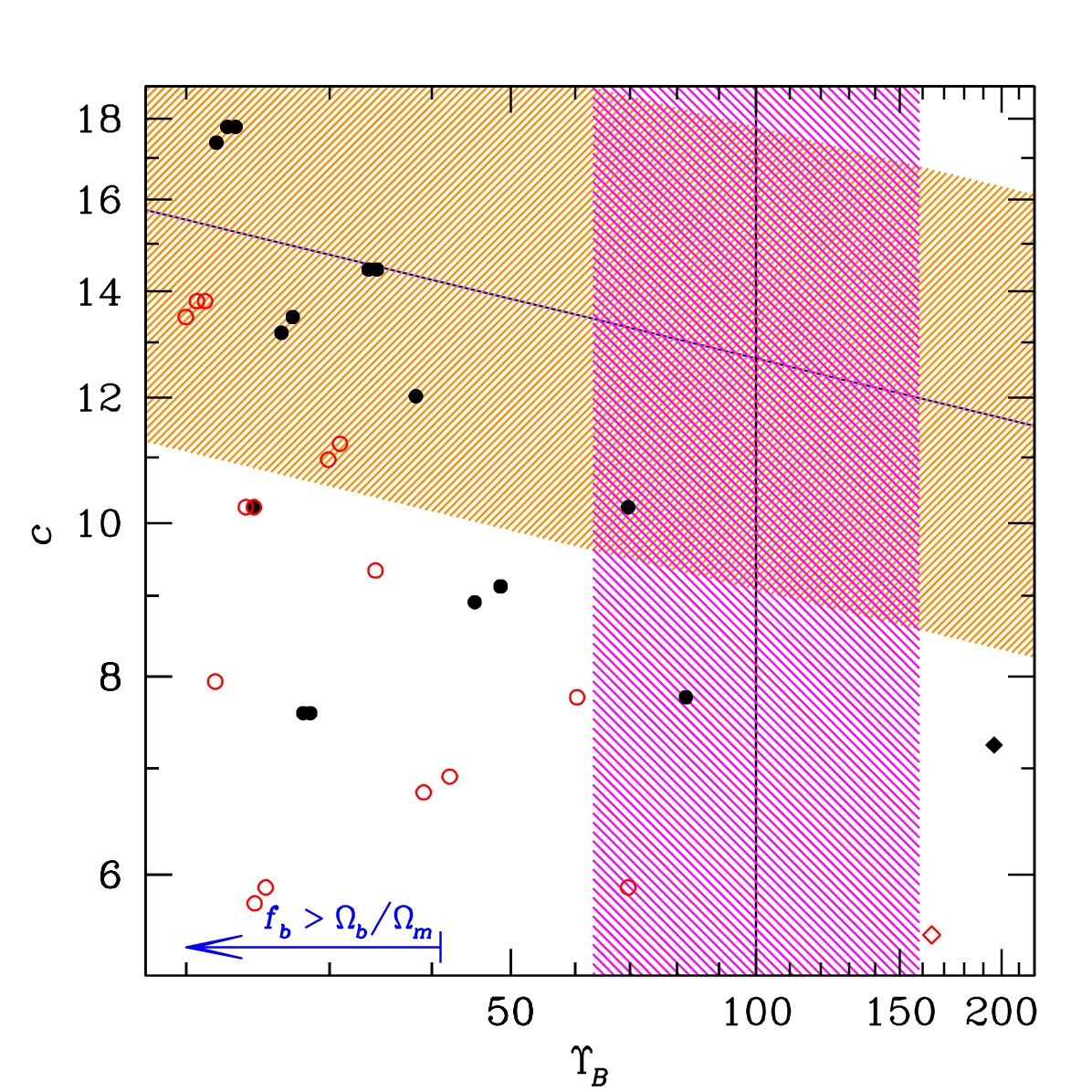}}
\caption{Dark matter (NFW) concentration versus total mass-to-light ratio 
at the virial radius,
$r_{200}$ (open circles) and $r_{102}$ (filled circles), as
inferred from the orbit solutions of 
Romanowsky et al. (2003,
supporting online material, \emph{circles}) and Romanowsky (2005, private
communication, \emph{diamonds}),
using equation~(\ref{xveq}).
The \emph{oblique horizontal line} indicates the concentration-mass that we
infer in paper~I from the dissipationless $\Lambda$CDM simulations of Navarro
et al. (2004), while the \emph{vertical line} gives the 
mass-to-light ratio at the virial radius from the cosmoslogical
prediction of Marinoni \& Hudson
(2002). The \emph{hashed regions} give an indication of the uncertainties on
the concentration-mass relation (factor 1.4) and the mass-to-light ratio at the
virial radius derived by Marinoni \& Hudson (factor 1.6).
The arrow indicates where the solutions lead to a baryonic fraction that is
higher than the universal value.
The filled symbols ought to lie within the hatched regions and to the right
of the arrow.
}
\label{chkRomUps}
\end{figure}

Figure~\ref{chkRomUps} shows the resulting values of the concentration
parameter of the NFW dark matter model versus the mass-to-light ratio
evaluated at the virial radii, $r_{200}$ (open circles) and $r_{102}$ (filled
circles), for \citeauthor{Romanowsky+03}'s 15 orbit solutions (with the
values of $r_s$, $v_*$, and $v_s^2/v_*^2$ given in their supporting
online material).  
Here, we first solved equation~(\ref{xveq}), and substituted the
dimensionless 
virial radius, $x_v$ for $x$ in equation~(\ref{MRom03ofr}).
We also plot (as diamonds) in
Fig.~\ref{chkRomUps} 
another acceptable solution, which A. Romanowsky kindly communicated to us, 
leading to $\Upsilon_B$ as high as 164 at $r_{200}$
and 196 
at $r_{102}$.

Whereas \citeauthor{Romanowsky+03} quote $\Upsilon_B = 33\pm14$ at 120 kpc,
we find, using the same input values of $r_s$, $v_*$, $v_s^2/v_*^2$, assuming
$R_e = 36''$,\footnote{The effective radius adopted by \cite{Romanowsky+03},
originates from aperture photometry \citep{RC3}.
Determinations from fits of the differential surface brightness profile
over a very large radial range
yield $R_e = 53''$ \citep{dVC79,CHLV90}, so the effective radius adopted by
\citeauthor{Romanowsky+03} may be underestimated by a factor 1.3.} 
(from which we infer a Hernquist scale radius of
$a = 1.0 \,\rm kpc$), 
$\Upsilon_B (120 \,\hbox{kpc})$ to be in the range $19 - 48$, with a
mean of 27 and a median of 22.
Extrapolating to the virial radius, defined in this paper as $r_{200}$,
we find $\Upsilon_B$ in the range 20 to 70, with a mean of 32 and a median
of 25, while for the $\Lambda$CDM virial radius, $r_{102}$, we get 
$\Upsilon_B$ between 22 and 82, with a mean of 37 and a median of 28.

Figure~\ref{chkRomUps} shows that 
\emph{one of the 15 orbit solutions of \citeauthor{Romanowsky+03} has a
combination of virial mass-to-light ratio and concentration parameter in line
with the predictions from dissipationless cosmological simulations}
($\Upsilon_{B,102} = 70$ and $c=10$), where 
the concentration parameter predictions are those that 
\cite{Napolitano+05}
rederived from the dissipationless
cosmological simulations analyzed by \cite{Bullock+01}. 
None of 
the orbit 
solutions lead to the very low concentration parameters favoured by
\cite{BSD03} and \citeauthor{Napolitano+05}.

We also note that the Hernquist model contribution to the mass-to-light ratio
at the virial radius is always near $\Upsilon_{*,B} = 5.6$.
According to equation~(\ref{gpos}), this amounts to a minimum mass-to-light
ratio of $\Upsilon_B = 41$ for the baryonic fraction at the virial radius to be
equal to the universal value. Only three of the fifteen orbit solutions of
\citeauthor{Romanowsky+03} can be modelled with no baryon bias, but for their
mean value of $\Upsilon_B = 33$, which we adopt as our lower limit,
the baryon bias is (eq.~[\ref{bbmin}])
$b_b = 1.25$ with $\Upsilon_{*,B} = 5.6$ or
$b_b=1.45$ with $\Upsilon_{*,B} = 6.5$.
Therefore,
\emph{in systems
with total mass-to-light ratios as low as the mean derived by
\citeauthor{Romanowsky+03}, the  
baryon fraction within the virial radius
is greater than the universal value}, unless the stellar mass-to-light ratio
is overestimated (see also \citealp{Napolitano+05}).
One is then left with the difficulty of explaining how baryons can
segregate from dark matter as far out as the virial radius.


\subsection{Velocity anisotropy}
\label{velanis}

Given the mass / anisotropy degeneracy mentioned in Sect.~\ref{intro}, it is
difficult 
to estimate the radial variation of the anisotropy parameter, $\beta$
(defined in eq.~[\ref{anisdef}]).
The simplest solution is to assume isotropy throughout the galaxy.
In fact, for clusters of galaxies, \cite{Merritt87}, \cite{LM03} and
\cite{KBM04} each derived 
isotropic orbits throughout, \citeauthor{Merritt87} and \citeauthor{KBM04}
by examining the global velocity distribution and \citeauthor{LM03} by
performing a joint local analysis of the radial profiles of the line-of-sight
velocity dispersion and kurtosis.
With their Schwarzschild-like \cite{GJSB98} modelling of elliptical galaxies, 
\cite{KSGB00} and \cite{SKGB00} find
$\beta \la 0.3$ within $R_e/2$.

\begin{figure}
\includegraphics[width=\hsize]{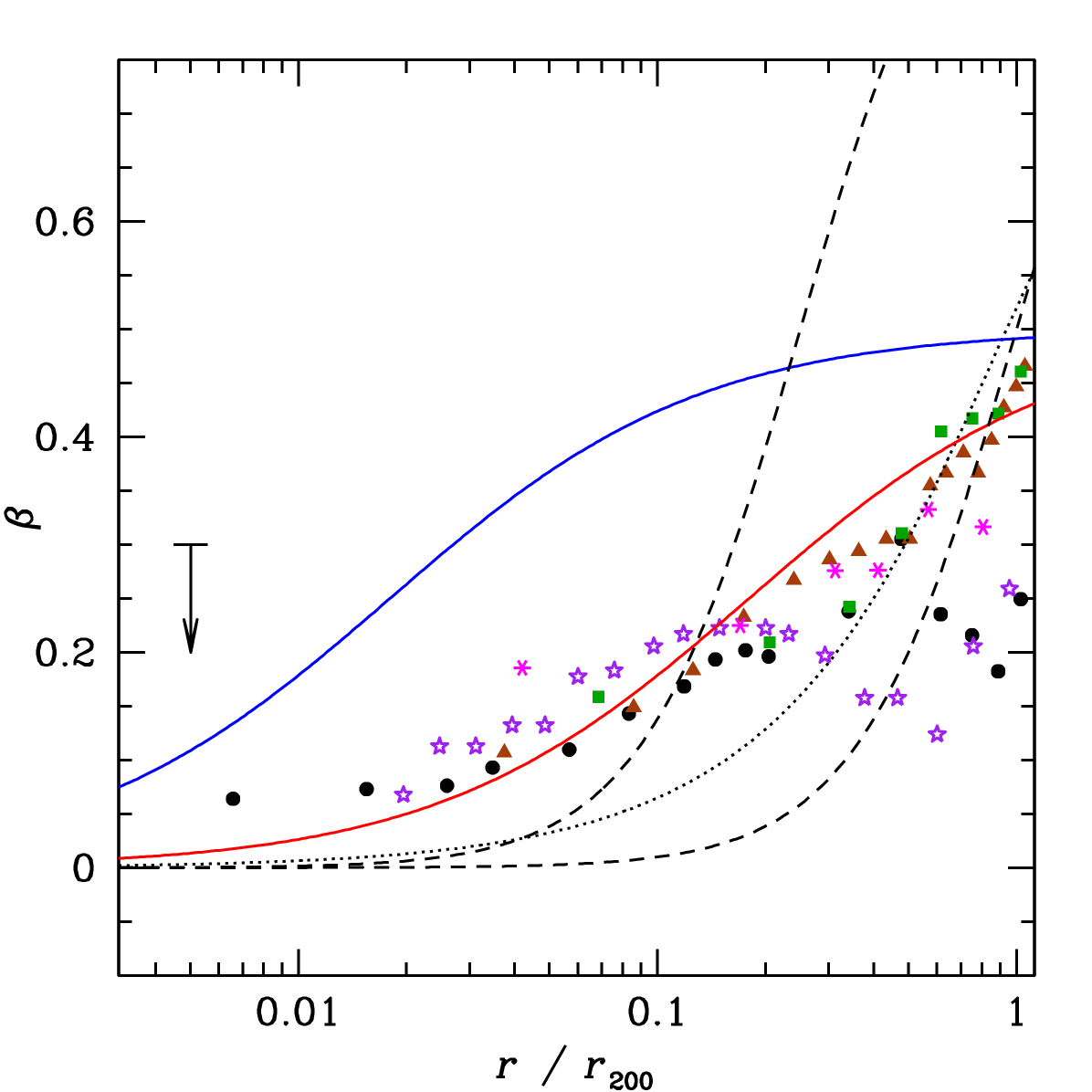}
\caption{Radial variation of the velocity anisotropy found in cosmological
$N$-body simulations.
The \emph{asterisks}, 
\emph{triangles}, \emph{stars}, 
\emph{circles}, and \emph{squares} correspond to particles within halos of
cosmological $\Lambda$CDM $N$-body simulations of
Diaferio (1999), Col\'{\i}n et al. (2000), Rasia et al. (2004),
Diemand et al. (2004b),
and Wojtak et al. (2005),
respectively.
The \emph{dotted curve} is the profile
proposed by 
Carlberg et al. (1997)
(eq.~[\ref{betaCarlberg}]) and found by 
Col\'{\i}n et al.
to fit well the
subhalo anisotropies.
%
The \emph{solid curves} are the analytical fit of equation~(\ref{betafit}),
with $r_a/r_v = 0.18$ (providing a good fit to the data from the
dissipationless cosmological $N$-body simulations) and 0.018 (matching the
anisotropy profile of Dekel et al.'s 2005 merger remnants), going upwards
(\emph{red} and \emph{blue} 
in the electronic version of the journal).
The \emph{dashed curves} are the Osipkov-Merritt anisotropy
(eq.~[\ref{betaOM}]) with $r_a/r_v =
0.25$ and 1, going downwards.
The upper limit is from the analysis of 
Kronawitter et al. (2000) and Saglia et al. (2000), assuming $r_v/R_e \simeq
100$ as expected for our standard parameters ($L_B = 10^{10}\, h_{70} {\rm
L_\odot}$ and $\Upsilon_B = 100$).
}
\label{betasims}
\end{figure}

It is useful to compare these still very uncertain anisotropy estimates with
those measured in cosmological $N$-body simulations.
Figure~\ref{betasims} shows the anisotropy profiles for five recent analyses of
dissipationless
cosmological simulations by 
\cite{Diaferio99}, \cite{CKK00}, \cite{RTM04}, \cite{DMS04_vel}, and
\cite{WLGM05}, 
 with the latter four converted from $r/r_{100}$ to
$r/r_{200}$ assuming $r_{100}/r_{200} = 1.37$, as expected for Nav04 models
with the masses (at the level of galaxy clusters) as simulated.
Note that the orbital
anisotropies of \cite{Diaferio99}, \cite{CKK00}, and \cite{WLGM05}
include the mean radial and tangential streaming motions, while that of
\cite{RTM04} does not.

The Figure also displays 
analytical representations of the data: the model 
 (\emph{dotted curve}) that
\cite{Carlberg+97_mofr} fit to the kinematics of CNOC clusters:
\begin{equation}
\beta(r) = 2\,\beta_m\,{r_a\,r\over r^2+r_a^2} \ ,
\label{betaCarlberg}
\end{equation}
with $\beta_m = 0.65$ and $r_a = 2\,r_v$, which \cite{CKK00} found to fit
well the anisotropy of the subhalos in their
simulation,
as well as an anisotropy model  that also appears to fit 
well the simulation data:
\begin{equation}
\beta(r) = {1\over2}\,{r\over r+r_a} \ ,
\label{betafit}
\end{equation}
for $r_a = 0.18$ (\emph{lower solid curve}).

The anisotropy profiles plotted in Figure~\ref{betasims} are
for all the particles of a halo, instead of for the subhalos of a halo, which
are much closer to isotropy.
Indeed, we are wary of using the subhalos in the cosmological simulations,  
because the number density profile  of subhalos
within halos is suspicious,
as it has a much shallower inner slope than the dark matter \citep{CKKK99}
with a nearly homogeneous core \citep{DMS04_rho},
in contrast with the
distribution of galaxies in clusters \citep{Carlberg+97_mofr}.
Interestingly, stars show more radial orbits than dark matter particles in
hydrodynamical
cosmological \citep{SDS04} and merger \citep{Dekel+05} simulations, where the
gas is allowed to cool and form stars, and, for the latter study, 
have a feedback effect on the
cooling of the remaining gas.
\citeauthor{Dekel+05} find
a \emph{stellar} anisotropy profile that
resembles the model of equation~(\ref{betafit}), with $r_a \simeq 1.4\,R_e
\simeq 0.018\,r_v$ (using eq.~[\ref{rviroverRe}] for the latter approximation).
This is shown as the higher of the two solid curves in Figure~\ref{betasims}.

Figure~\ref{betasims} indicates that the anisotropy of
equation~(\ref{betafit}) with $r_a = 0.18\,r_v$ 
fits very well the anisotropy profiles of \citeauthor{CKK00},
\citeauthor{RTM04}, and
\cite{DMS04_vel} 
from 2\% to 10\% of the 
virial radius, which happens to be the region where we shall find that the
anisotropy may play an important role in modelling the mass-to-light ratios
of elliptical galaxies.
Figure~\ref{betasims} also shows the
commonly-used
Osipkov-Merritt \citep{Osipkov79,Merritt85_df} anisotropy
model, 
\begin{equation}
\beta(r) = {r^2 \over r^2+r_a^2 } \ ,
\label{betaOM}
\end{equation}
for $r_a/r_v = 0.25$ and 1.
Clearly, \emph{the Osipkov-Merritt anisotropy 
is a poor fit to the simulations}, as it converges
to a too high value of unity, and worse, 
decreases to zero too fast at increasingly 
low radii.

In what follows, we generally assume
that the anisotropy of the stellar population
is equal to the anisotropy of the dark matter particle system, but we will
also allow for the much stronger radial anisotropy found by
\citeauthor{Dekel+05}.
In appendix~\ref{appJeans}, we derive the line-of-sight velocity dispersion
profile for four different anisotropy profiles: constant anisotropy, the
limiting purely radial case, Osipkov-Merritt,
and the anisotropy model of equation~(\ref{betafit}).
With these single quadratures, one avoids double integration (first
integrating the Jeans 
equation to obtain the radial velocity dispersion, and then integrating along 
the line of sight to get the line-of-sight velocity dispersion). 

\section{Results}
\label{darkmass}

If, as we found in paper~I,
the mass models found in dissipationless cosmological $N$-body simulations 
are not able to reproduce by themselves the rather
high central velocity dispersions observed in elliptical galaxies,
this suggests that the central velocity dispersions of ellipticals are dominated
by the stellar component and by a central supermassive
black hole, as we shall see below.
We are now left with the question of whether 
the NFW, JS--1.5 and Nav04 models are adequate in describing
the diffuse dark matter component (excluding the central black hole) 
of elliptical galaxies.

Dark matter is expected to become significant in the outer
regions of ellipticals, since the black hole affects the central regions, and
the influence of the stellar component is usually thought to be
important in the inner regions, at least within $R_e$.
We therefore predict the line-of-sight velocity dispersion profiles of
ellipticals built with 4 components: 
a S\'ersic stellar component of constant mass-to-light ratio
$\Upsilon_*$, a central black hole, a hot gas component,
and a diffuse dark matter component described by an NFW,
JS--1.5, or Nav04 model, according to our parameterisations of Sec.~\ref{basic}.
We focus our analysis on the following set of parameters: a luminosity $L_B =
L_{*,B} = 1.88\times 10^{10}\,h_{70}^{-2}\,{\rm L_\odot}$ (implying 
$R_e = 3.2 \, h_{70}^{-1} \, \rm
kpc$ from eq.~[\ref{Refit}], $m=3.15$ from eq.~[\ref{mfit}], and $a_S = 12 \,
h_{70}^{-1} \, \rm pc$ from eq.~[\ref{aSfit}]), 
a stellar mass-to-light ratio
$\Upsilon_{*,B} = 6.5$ and no baryon bias ($b_b=1$).

\subsection{The effects of velocity anisotropy}
\label{aniseffects}

\begin{figure}
\includegraphics[width=\hsize]{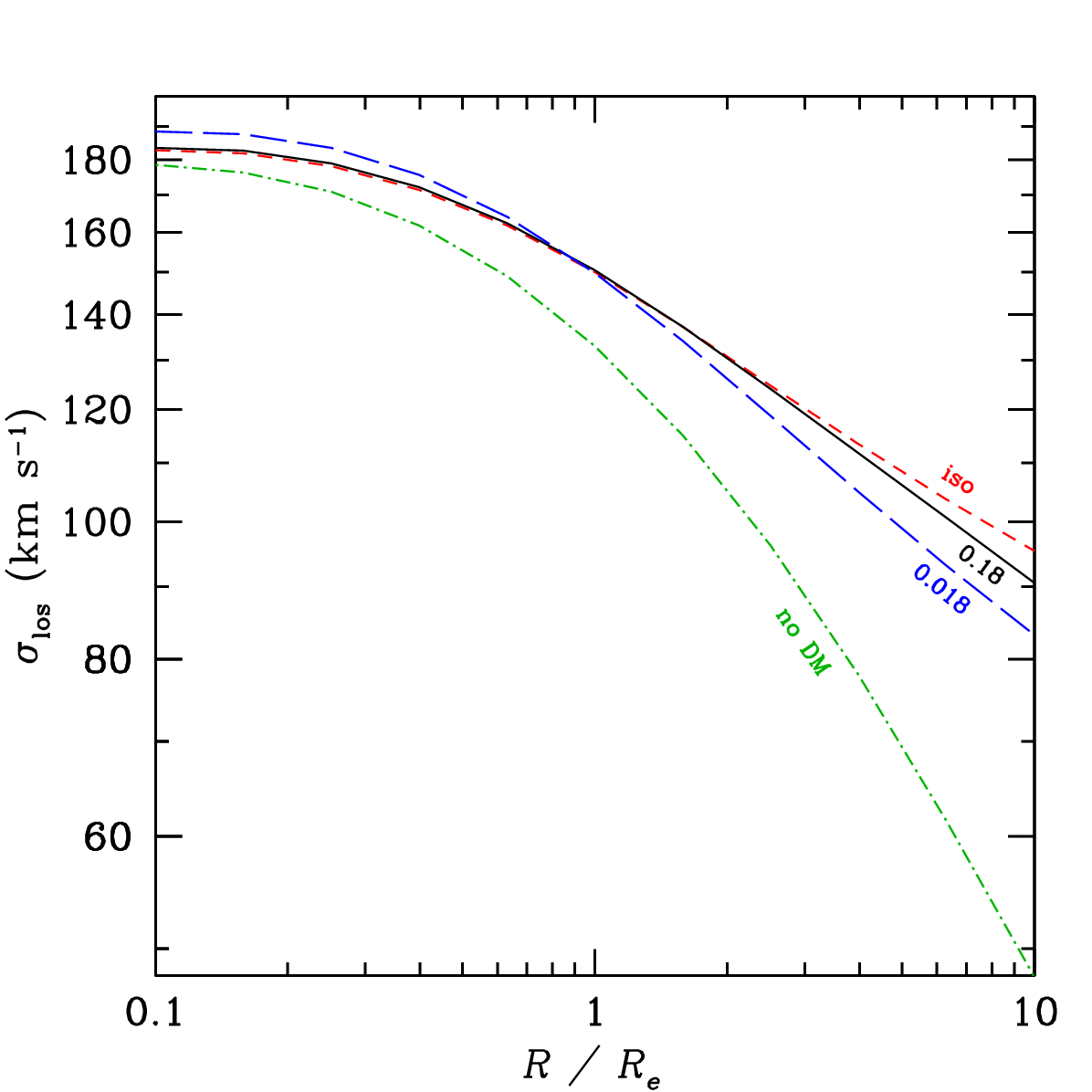}
\caption{Line-of-sight stellar 
velocity dispersion of 4-component elliptical galaxy 
for different anisotropy models
(eqs.~[\ref{sigall}] and [\ref{sigallker}]), 
with a Nav04 dark matter model,
$L_B = L_{*,B} = 1.88 \times 
10^{10} {\rm L_\odot}$, $m=3.15$, $R_e = 3.2 \, h_{70}^{-1} \, \rm kpc$,
$\Upsilon_{*,B} = 6.5$,
$\Upsilon_B = 100$,
$g_{\rm BH} = 0.0015$ and $b_b = 1$.
The \emph{short-dashed}, \emph{solid}, and \emph{long-dashed curves}
represent the 
isotropic (see eq.~[\ref{siglosiso}]), mildly and sharply 
anisotropic (eq.~[\ref{betafit}], with $r_a/r_v = 0.18$ and 0.018), using
equation~[\ref{siglosnewbeta}]), 
respectively, while the \emph{dash-dotted curve} represents a galaxy with no
dark matter ($\Upsilon_B = f_b\,\Upsilon_B^{\rm std} = 14$)
and mild anisotropy (eq.~[\ref{betafit}] with
$r_a/r_v = 0.18$) --- these four cases are
respectively shown in \emph{red}, \emph{black}, \emph{blue} and \emph{green}
in the electronic version of the journal.
}
\label{siglosanis}
\end{figure}
We first check the effects of anisotropy, by computing the line-of-sight
velocity dispersions of our four-component model, assuming different anisotropy
profiles.
Figure~\ref{siglosanis} shows the total
line-of-sight velocity dispersion (i.e., the quadratic sum of the individual
velocity dispersions of the four
components) for four
anisotropy models: isotropic, radial, and the model of
equation~(\ref{betafit}) for two choices of $r_a$.
The line-of-sight velocity dispersion profile 
of the anisotropy model of equation~(\ref{betafit}) with
our standard value of $r_a = 0.18\,r_v$ is virtually
indistinguishable from that of the isotropic model: the anisotropic model
producing 2\% lower velocity dispersions at $5\,R_e$.
On the other hand, if the velocity ellipsoid is indeed radially 
anisotropic beyond
$r_v/50$, as suggested by the orbital anisotropies of the 
stellar particles in the simulations of \cite{SDS04} and \cite{Dekel+05},
then one can obtain a
small but non-negligible decrement in line-of-sight velocity dispersion: if
$r_a = 0.018\,r_v$, one finds decrements of 4\% at $2\,R_e$ and 9\% at
$5\,R_e$.
Note that with Osipkov-Merritt anisotropy, and $r_a = r_v$ or even $r_a =
0.25\,r_v$, the
line-of-sight velocity dispersions (obtained with eq.~[\ref{siglosOM}])
are indistinguishable from the isotropic ones.

Hence, with an optimistic precision on velocity dispersion of $5\,\, \rm km
\, s^{-1}$, \emph{the slight radial anisotropy found in cosmological $N$-body
simulations produces line-of-sight velocity dispersion profiles that are
virtually indistinguishable within $9\,R_e$ 
from analogous profiles assuming velocity
isotropy everywhere, 
while the stronger radial anisotropy found for stars
in simulated merger remnants displays measurable differences beyond 
$2\,R_e$}.

In comparison, as illustrated in Figure~\ref{siglosanis}, 
discarding the dark matter component has a much stronger effect on the velocity
dispersions at a few effective radii
than does the strong radial anisotropy with $r_a = 0.018\,r_v$, with
dispersion 19\% and 33\% lower than with dark matter (at $\Upsilon_B=100$).

\subsection{Relative importance of dark matter, stars, gas and the central
  black hole}
\label{components}

\begin{figure*}
\includegraphics[width=\hsize]{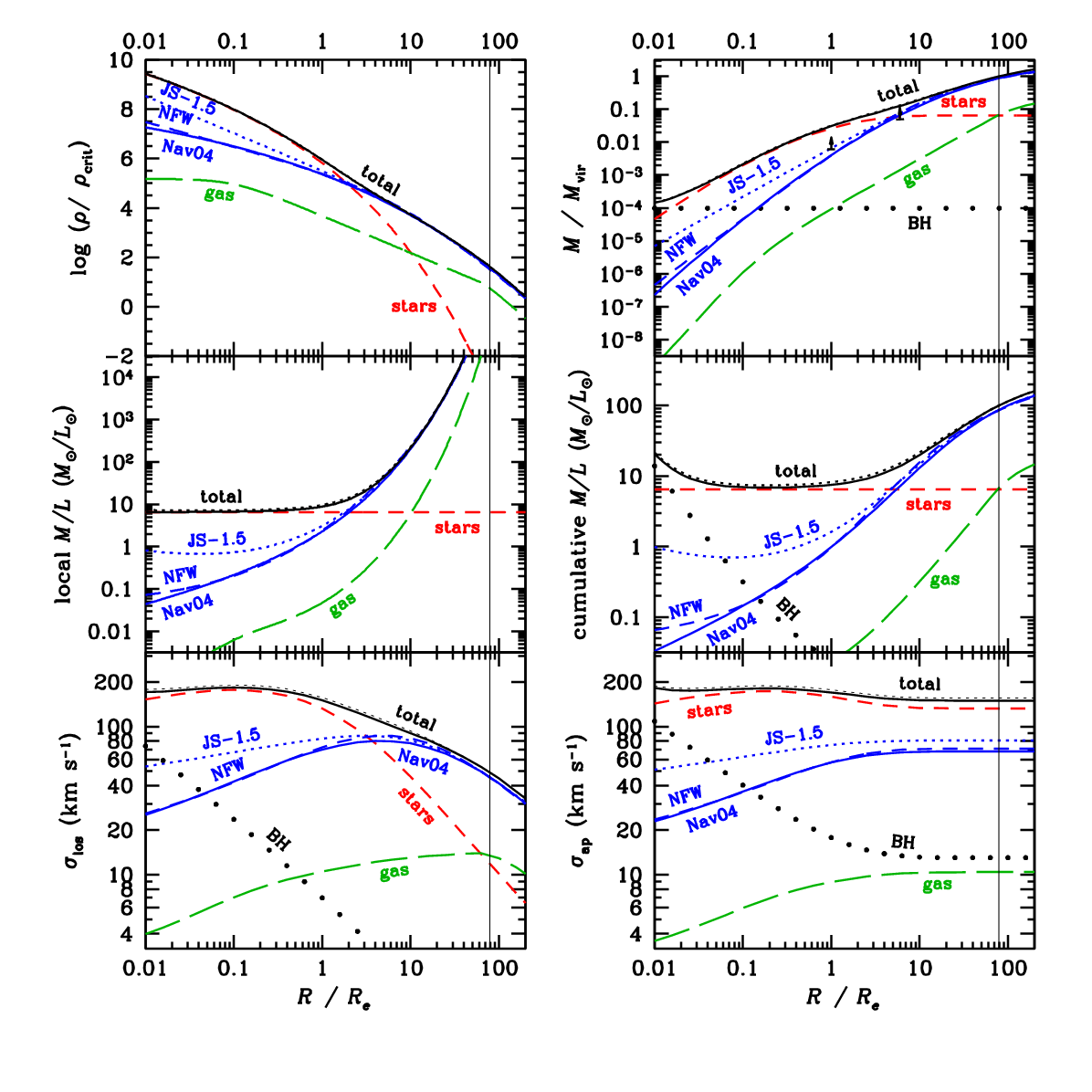}
\caption{Contribution of stars (\emph{short dashes}), dark matter
  (\emph{light curves};  \emph{dashed} for NFW, \emph{dotted} for JS--1.5,
and \emph{solid} for Nav04), hot 
  gas (\emph{long dashes}), and the central black hole (\emph{big dots})
to the local mass density (\emph{upper left}),
 cumulative mass (\emph{upper right}),
local (\emph{middle left}) and cumulative (\emph{middle right}) mass-to-light
ratios, 
mildly anisotropic (eq.~[\ref {betafit}] with $r_a = 0.18\,r_v$)
line-of-sight velocity dispersion (\emph{lower left}) and isotropic 
aperture velocity dispersion (\emph{lower right});
for a giant elliptical with $L_B = L_{*,B} = 1.88 \times 
10^{10} h_{70}^{-2} {\rm L_\odot}$
  (corresponding to $m=3.12$, $R_e=3.2 \, h_{70}^{-1} \, \rm kpc$ and
$a_S=11.6 \, h_{70}^{-1} \, \rm pc$), $\Upsilon_{*,B}=6.5$, $\Upsilon_B = 100$
(i.e., $b_{\Upsilon,B} = 0.26$), 
yielding $r_v/R_e = 79$, shown as the \emph{thin
  vertical lines}) with no baryon bias
   ($b_b=1$), black hole to stellar mass ratio $g_{\rm BH} =
0.0015$ and dark matter scaling relations from
  Sec.~\ref{basic}. 
The \emph{top, solid} and \emph{dotted, curves} represent
the total galaxy assuming either Nav04 or JS--1.5 dark matter,
respectively. 
The \emph{arrows (upper right)} 
indicate the lower limits to the dark matter mass according to
Loewenstein \& White (1999), given our total mass at those radii.
}
\label{figcomponents}
\end{figure*}
Figure~\ref{figcomponents} shows the contribution of each component to 
the local mass density, cumulative mass, line-of-sight (with the mild
anisotropy of eq.~[\ref{betafit}] and $r_a/r_v=0.18$)  and aperture
(isotropic\footnote{The aperture velocity  dispersion with the anisotropy of
eq.~[\ref{betafit}] is difficult to express in terms of a single quadrature,
but the difference with the isotropic case should
be very small, since the difference between isotropic and mildly anisotropic
line-of-sight velocity dispersions is negligible, see Fig.~\ref{siglosanis}.})
velocity
dispersions.
For our standard parameters, $L_B = L_{*,B} = 1.88 \times
10^{10}\,h_{70}^{-2} {\rm L_\odot}$, 
$\Upsilon_{*,B}=6.5$, $\Upsilon_B = 100$   
(the typical value inferred in the cosmological analyses of \citealp{MH02}
  and \citealp{YMvdB03}, i.e. $b_\Upsilon=0.26$), and $g_{\rm BH} = 0.0015$,
\emph{the stellar
component dominates over the dark matter component out to 
$R \simeq 2\,R_e$ for the local density, $5\,R_e$ for the enclosed mass, 
$3\,R_e$
for the line-of-sight velocity dispersion, while the stars dominate the dark
matter everywhere for the aperture velocity dispersion}.
In the inner regions, with $g_{\rm BH} = 0.0015$,
\emph{the central black hole is the dominant component for $R
< 0.015\,R_e$ for the enclosed mass, $0.006\,R_e$ for the aperture velocity
dispersion, but only out to $0.003\,R_e$ (i.e., $\simeq 10\,\rm pc$) for the
line-of-sight velocity 
dispersion}. 

The figure
also indicates that our model 
produces Nav04 dark matter contents that at $R=5\,R_e$ are consistent
with
the joint kinematics / X-ray modelling of \cite{LW99}, while our Nav04
dark matter
mass fraction at $R=R_e$ is 1.6 times lower than the lower limit found by
\citeauthor{LW99}
(but our predicted JS--1.5 dark matter mass is precisely the lower limit of
\citeauthor{LW99}). 

Although the gas component begins to dominate the stellar component at
$10\,R_e$, as expected, the gas does not influence the line-of-sight or
aperture velocity dispersions: although its influence on the line-of-sight
velocity dispersion dominates that of the stellar component at $60\,R_e$, by
that large radius the dark matter component is fully dominating the 
stellar velocity dispersion.

Given the strong dominance of stars (and the central black hole) over the dark
matter component, \emph{the differences in aperture or line-of-sight
velocity dispersion profiles between the inner dark matter slopes of $\alpha
= 1$ and $\alpha = 3/2$ are very small, and basically indistinguishable from
observations}. 

\begin{figure}
\includegraphics[width=\hsize]{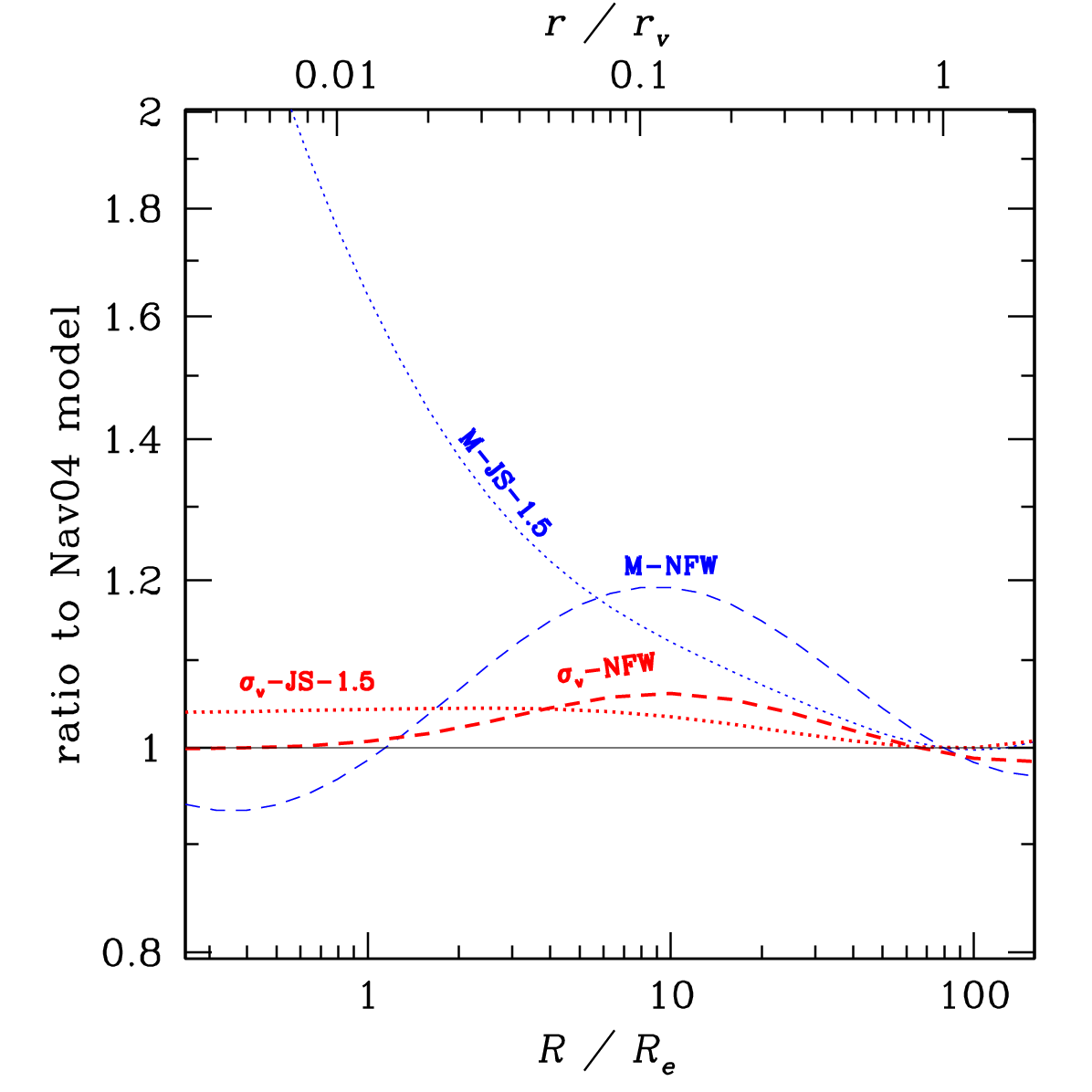}
\caption{Radial profiles of cumulative dark mass (\emph{thin curves},
\emph{blue} in the electronic version of the Journal) and
  stellar line-of-sight velocity dispersion
(\emph{thick curves}, \emph{red} in the electronic version of the Journal)
for the NFW 
  (\emph{dashed curves}) and JS--1.5 (\emph{dotted curves}) dark matter
  models, both normalized to those of the Nav04 dark matter model.
Same parameters as in Figure~\ref{figcomponents}.
}
\label{ratios}
\end{figure}

The line-of-sight velocity dispersion using the Nav04 profile is also
indistinguishable from the other two dark matter profiles. However, the Nav04
profile produces slightly ($5\%$) lower total velocity dispersions at 
$R=5\,R_e$.
This is illustrated in Figure~\ref{ratios}, which plots the mass and
line-of-sight velocity dispersion profiles normalized to those of the Nav04
model.
At radii around $r_v/8$ (corresponding to $10\,R_e$ for our standard set of
parameters), the NFW velocity dispersion is up to 6\% larger than 
that of the Nav04 model, which fits much better the density
profiles found in dissipationless 
cosmological $N$-body simulations. The corresponding
overestimate of the cumulative mass is almost 20\% at $r_v/8$ ($10\,R_e$),
and the same effect is visible in Figure 2 of \cite{Navarro+04} 
(with the virial radii given in their Table 3).
%
The maximum velocity dispersion ratio in Figure~\ref{ratios} indicates that
the lower stellar velocity dispersions obtained with the Nav04 dark matter
model relative to its NFW counterpart are not caused by the convergent dark
matter mass
profile of the former at large radii, but by the 20\% difference in mass
profiles at 
$10\,R_e$.

The conclusions from Figure~\ref{figcomponents} are unchanged if we adopt the
much higher universal 
total mass-to-light ratio $\Upsilon_B = \overline \Upsilon_B = 390$,
instead of 100, in particular the stellar component dominates the dark matter
out to only $2\,R_e$ (instead of $3\,R_e$)
for the line-of-sight velocity dispersion.


Note that the ratio of aperture velocity dispersion at $R_e/8$ (the aperture
used by \citealp{JFK96}) to
circular velocity at the virial radius, $\sigma_{\rm ap}(R_e/10)/V_v$,
is 0.64 when $\Upsilon_B = 390$, but as
high as 1.00 when $\Upsilon_B = 100$. Indeed, since the stars dominate the
aperture 
velocity dispersion measurement, decreasing the total mass-to-light ratio
decreases $V_v$ without affecting much $\sigma_{\rm ap}(R_e/8)$.

\begin{figure}
\resizebox{8.5cm}{!}{\includegraphics{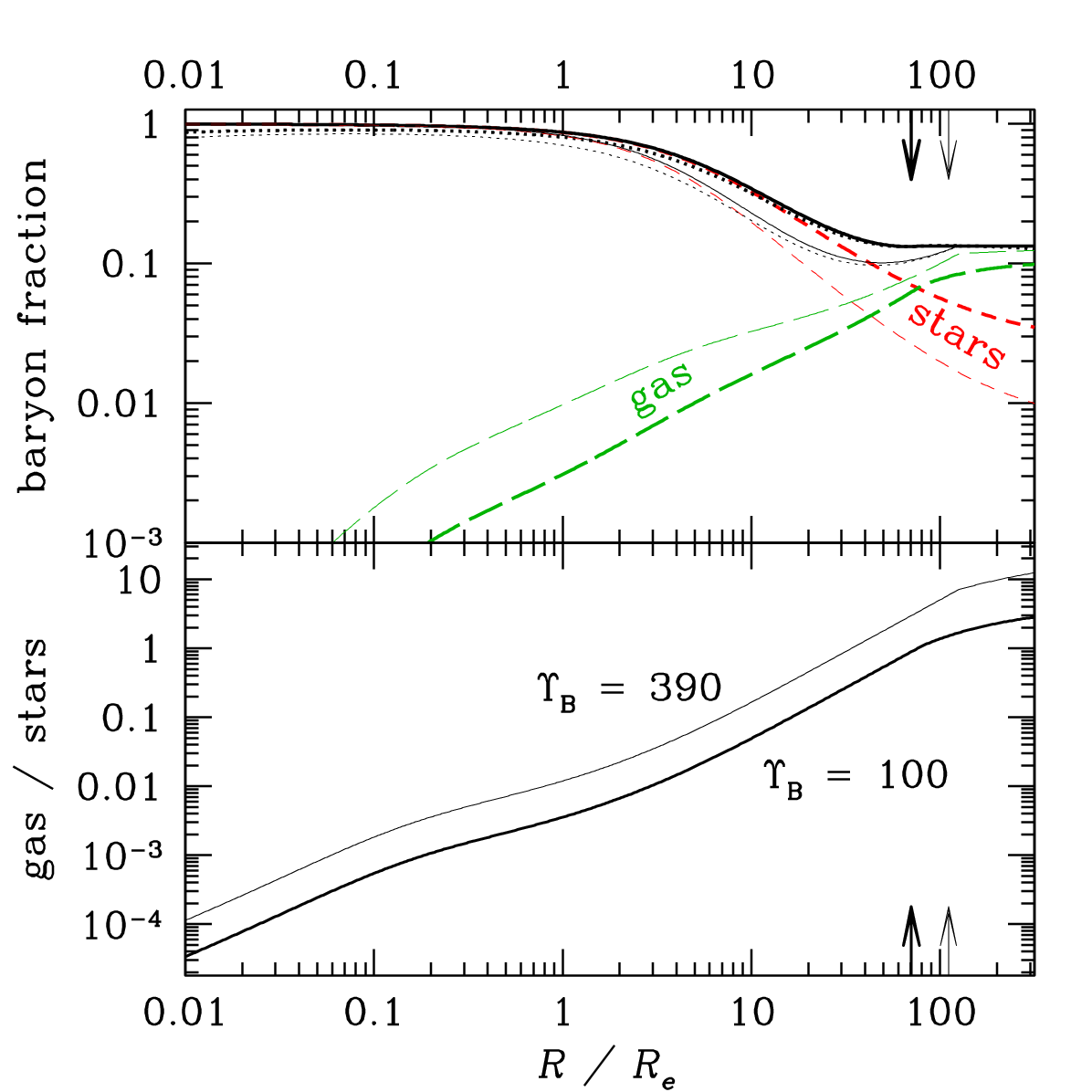}}
\caption{Radial profiles of cumulative fractions of
gas (\emph{long dashes}), stars (\emph{short
dashes}) and total baryons (\emph{solid curve}) for a Nav04 dark
matter model (\emph{top}), and the cumulative gas/star ratio (\emph{bottom}).
Also shown is the baryon fraction with the JS--1.5 model (\emph{dotted} in
\emph{top} panel).
The \emph{arrows} indicate the virial radii.
\emph{Thick} and \emph{thin curves} and \emph{arrows} are for $\Upsilon_B =
100$ and 390, 
respectively.
Other parameters are the same as in Figure~\ref{figcomponents}.
}
\label{gsfracs}
\end{figure}

Figure~\ref{gsfracs} displays radial profiles of the baryon fraction and
gas-to-star ratio.
For our standard mass-to-light ratio at the virial radius, $\Upsilon_B =
100$,
the decrease in star fraction with radius is only partially compensated by
the increase in gas fraction, so that \emph{the total baryon fraction
decreases with radius for $r < r_v$}: for the Nav04 model,
$f_b$ decreases from 
unity at small radii to 0.87 at $R_e$, 0.76 at $2\,R_e$, 0.53
at $5\,R_e$, and 
down to a minimum of 0.132 at $R = 68\,R_e \simeq 0.95\,r_v$, and
then slowly rises by 0.1\% from there to the virial radius. 
These trends are very similar for the NFW dark matter model, while for the
JS--1.5 dark matter 
model, the baryon fraction actually increases from
0.87 at $R_e/100$ to 0.90 at $R_e/13$, then decreases to 0.80 at $R_e$, 0.70
at $2\,R_e$, 0.49 at $5\,R_e$.
For the universal mass-to-light ratio, $\Upsilon_B = 390$, the trends are
similar, except that the baryonic fraction reaches a minimum of 0.10 
(for all dark matter models) around $50\,R_e$, i.e, $r_v/2$,
and rises at larger radii.

The bottom plot of Figure~\ref{gsfracs} indicates that the gas-to-star ratio
rises, 
roughly as $r^{1.2}$ from $r_v/10\,000$ to $r_v$.
If the mass-to-light ratio at the virial radius is increased, while the
baryonic fraction at the virial radius and the total stellar mass are kept
constant, then the gas component must become relatively 
more important, which explains
why the $\Upsilon_B = 390$ gas-to-star ratio is larger than the $\Upsilon_B =
100$ gas-to-star ratio, by a factor $390/100 = 3.9$.


\subsection{Can one weigh the dark matter component?}
\label{weighdm}

As we have seen in paper~I and
Sect.~\ref{components}, the dark
matter contributes little to the inner regions of elliptical galaxies, and
hence we need to focus on the outer regions to be able to weigh the dark
component.

\begin{figure}
\includegraphics[width=\hsize]{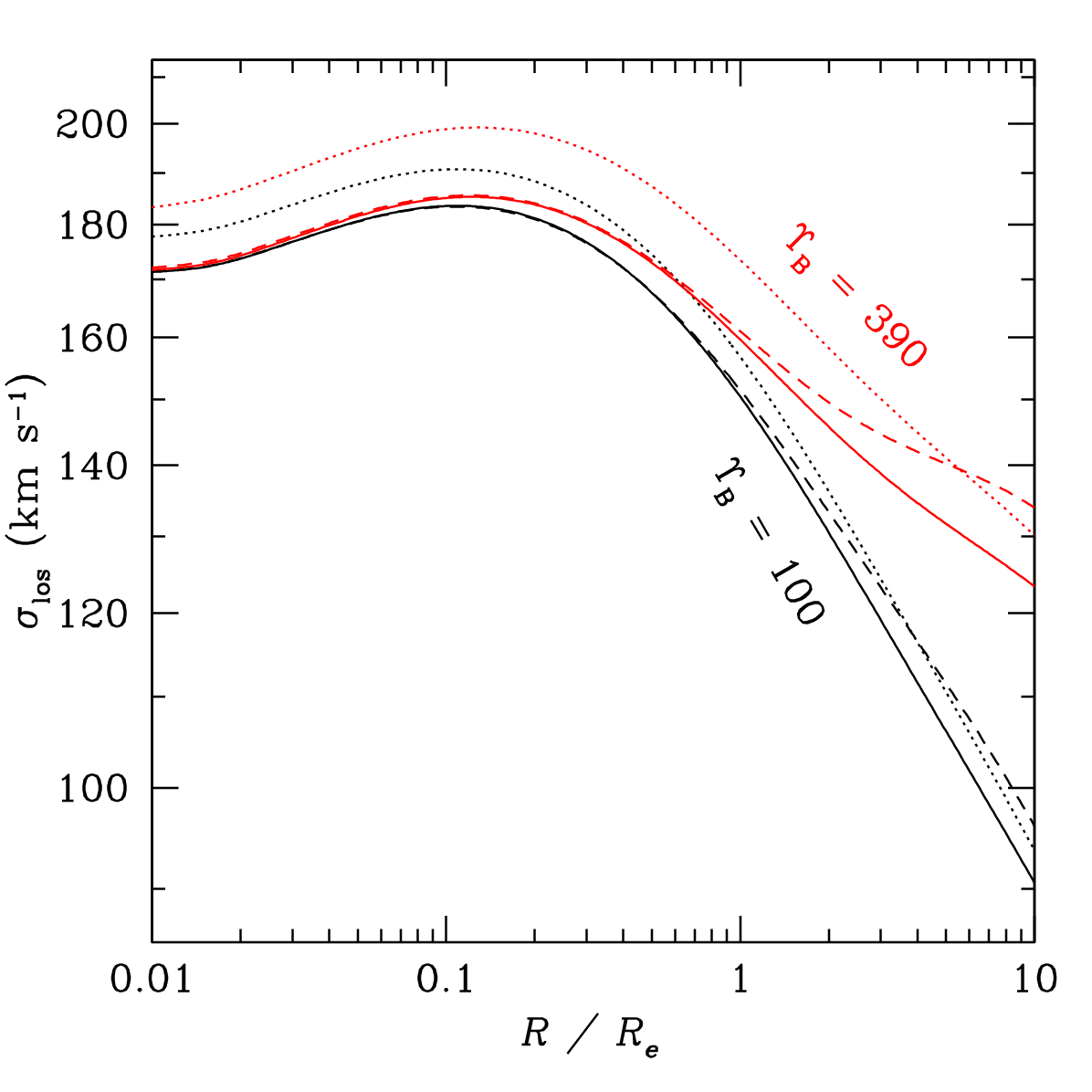}
\caption{Line-of-sight  stellar 
velocity dispersions of our 4-component elliptical galaxy 
(eq.~[\ref{siglosnewbeta}]) for our slightly anisotropic
(eq.~[\ref{betafit}] with $r_a = 0.18\,r_v$)
NFW (\emph{dashed}), JS--1.5
(\emph{dotted}) and Nav04 (\emph{solid}) dark matter models, with 
total mass-to-light ratio within the virial radius 
$\Upsilon_B
= 100$ and 390, increasing upwards (respectively
\emph{black} and \emph{red} in the electronic version of the Journal).
Same other parameters as Figure~\ref{figcomponents}.
}
\label{siglosM}
\end{figure}
Figure~\ref{siglosM} illustrates the effect of the mass of the dark matter
component on the line-of-sight velocity dispersion profiles.
While the steep inner slope of the density profile of the JS--1.5 model
allows it to have a non-negligible effect on the line-of-sight velocity
dispersions at small projected radii, the opposite is true for the NFW and
Nav04 models: the inner line-of-sight velocity dispersions are completely
independent of the total mass of the galaxy.
Moreover, at large radii with $\Upsilon_B=100$, it is difficult to
distinguish between the three dark matter models.
%
Note that relative to the Nav04 dark matter model,
the NFW  model has no effect on the line-of-sight velocity
dispersions measured from 0.01 to $0.1\,R_e$ (even when no central black hole
is present), while the JS--1.5 component has a
small effect, increasing with mass. 

\begin{figure}
\includegraphics[width=\hsize]{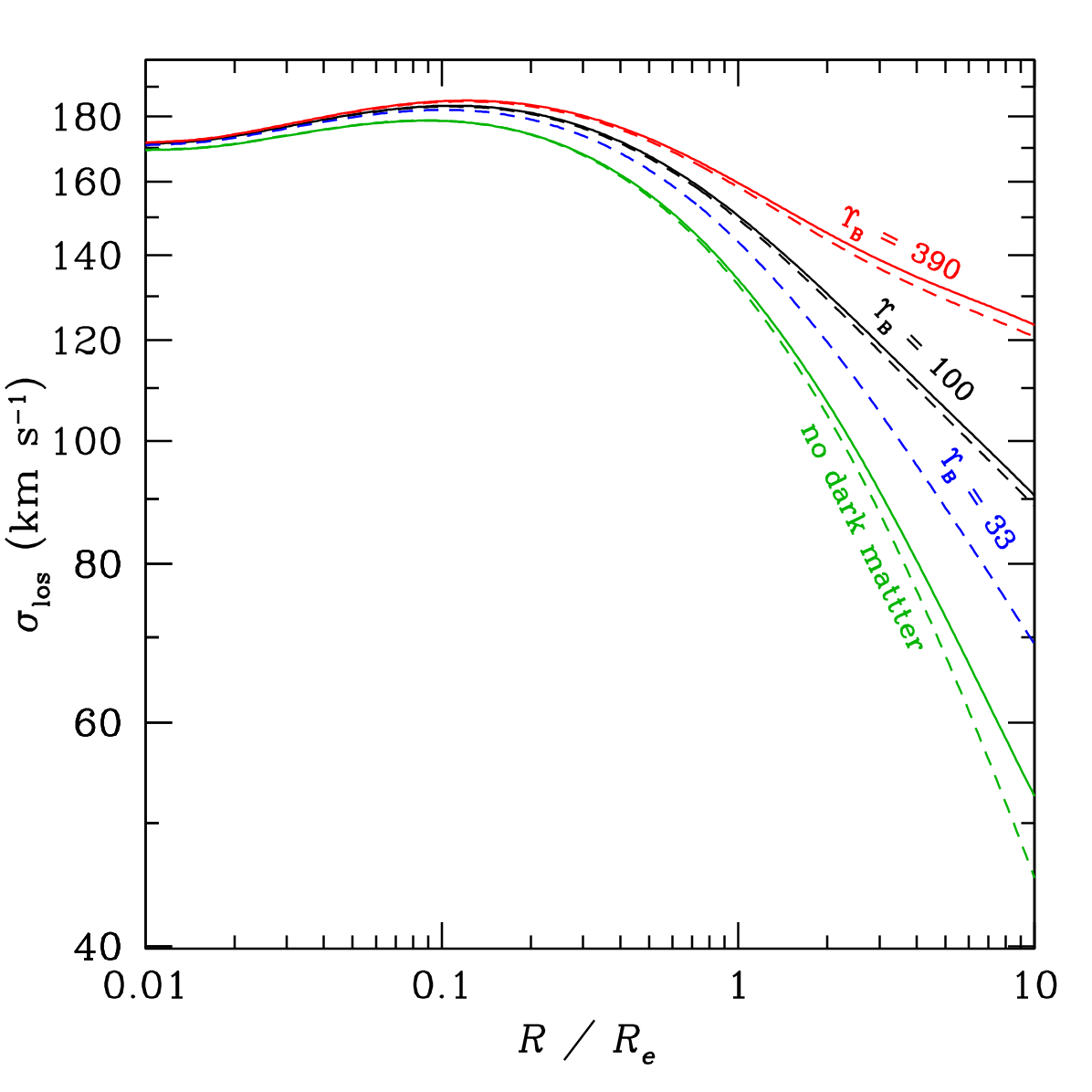}
\caption{Same as Figure~\ref{siglosM} with Nav04 dark matter,
$\Upsilon_B = 33$, 100, and 390, increasing upwards (\emph{blue}, \emph{black},
and 
\emph{red} in the electronic version of the Journal), as well as no dark
matter ($\Upsilon_B = f_b\,\Upsilon_B^{\rm std} = 14$)
case (\emph{lower curve}, \emph{green} in the electronic version of
the Journal), for which $\Upsilon_B=f_b\,\Upsilon_B^{\rm std} = 14$.
The baryon bias is $b_b=1$
(\emph{solid curves}) and $b_b = 1.5$ (\emph{dashed curves}).
The case $b_b=1$ is not shown for $\Upsilon_B=33$ because the gas fraction
would then be negative (see eq.~[\ref{gpos}]).
Same other parameters as in Figure~\ref{figcomponents}.
}
\label{siglosMbb}
\end{figure}

Figure~\ref{siglosMbb} shows again the effect of the mass of the dark matter
component, this time allowing the very low mass-to-light ratio deduced by
\cite{Romanowsky+03} (see Sect.~\ref{UpsRomsec}), 
which requires $b_b>1$ (eq.~[\ref{bbmin}]).
%
The figure 
shows that,
at $R/R_e = 2$, 3, 4, 5, and 6, the velocity dispersion increases by 
26, 33, 39, 43, and $46 \, \rm km \, s^{-1}$ when the mass-to-light ratio at
the virial radius is increased from $\Upsilon_B = 33$ to $\Upsilon_B = 390$.
In other words,
\emph{a line-of-sight 
velocity dispersion measurement at $4\,R_e$ with $20 \, \rm km
\, s^{-1}$ 
measurement error 
implies an uncertainty on the total
mass-to-light ratio of a factor greater than 3}!
At higher radii, less precision is required on the velocity dispersions, but
these are more difficult to estimate as they are smaller.
These conclusions are the same if we adopt purely isotropic models.
This three-fold uncertainty becomes even larger if we do not assume a
precise form for the velocity anisotropy profile.

Also shown in Figure~\ref{siglosMbb} is the case of no dark matter (for which
$\Upsilon_B = f_b\,\Upsilon_B^{\rm std} = 14$). The dispersions now fall to
quite low values. However, this `no dark matter' case is extreme, as its
baryonic fraction is unity.

\begin{figure}
\includegraphics[width=\hsize]{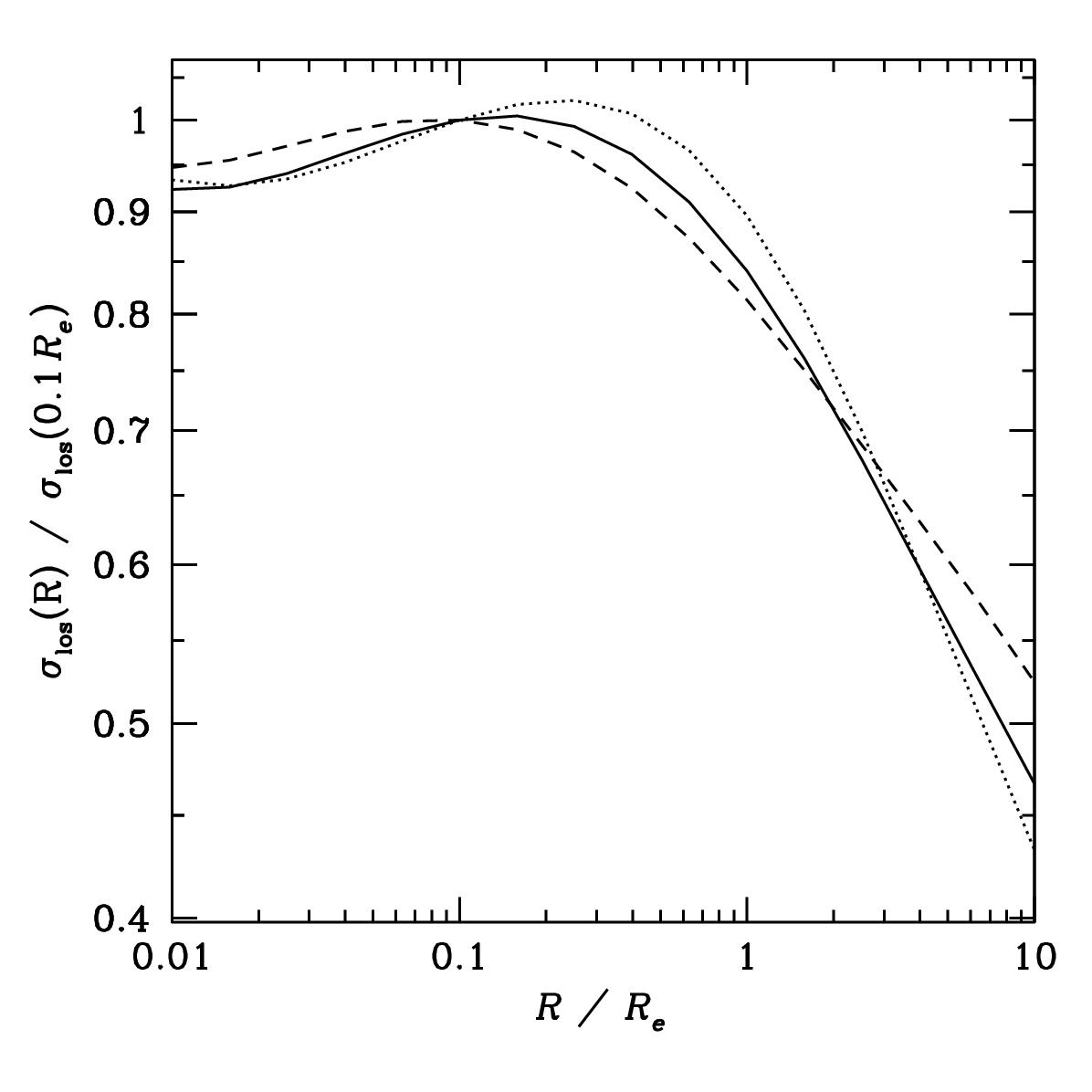}
\caption{Normalized 
line-of-sight stellar velocity dispersion for our slightly anisotropic
(eq.~[\ref{betafit}] with $r_a = 0.18\,r_v$) model with Nav04 dark matter,
with $\Upsilon_B = 100$ and $L_B = 3.2\times10^9$ (\emph{dotted curve}),
$10^{10}$ (\emph{solid curve}), and $3.2\times 10^{10} {\rm L_\odot}$ (\emph{dashed
curve}), with S\'ersic shape $m$ and effective radius $R_e$ taken from
equations~(\ref{mfit}) and (\ref{Refit}), respectively, and 
other parameters as in Figure~\ref{figcomponents}.}
\label{siglosL}
\end{figure}

How much do these conclusions depend on the galaxy luminosity?
Figure~\ref{siglosL} displays the dispersion profiles normalized to the
values at $0.1\,R_e$ (at which radius the dispersion profile is fully
dominated by the stellar component, as seen in Fig.~\ref{figcomponents}).
The velocity dispersion profiles have a strikingly similar shape, with
relative differences less than 6.5\% for $R < 5\,R_e$, and interestingly, less
than 3.5\% for $R \simeq 2.5\,R_e$.
Also, the slope at $R > R_e$ varies with luminosity.

\begin{figure}
\includegraphics[width=\hsize]{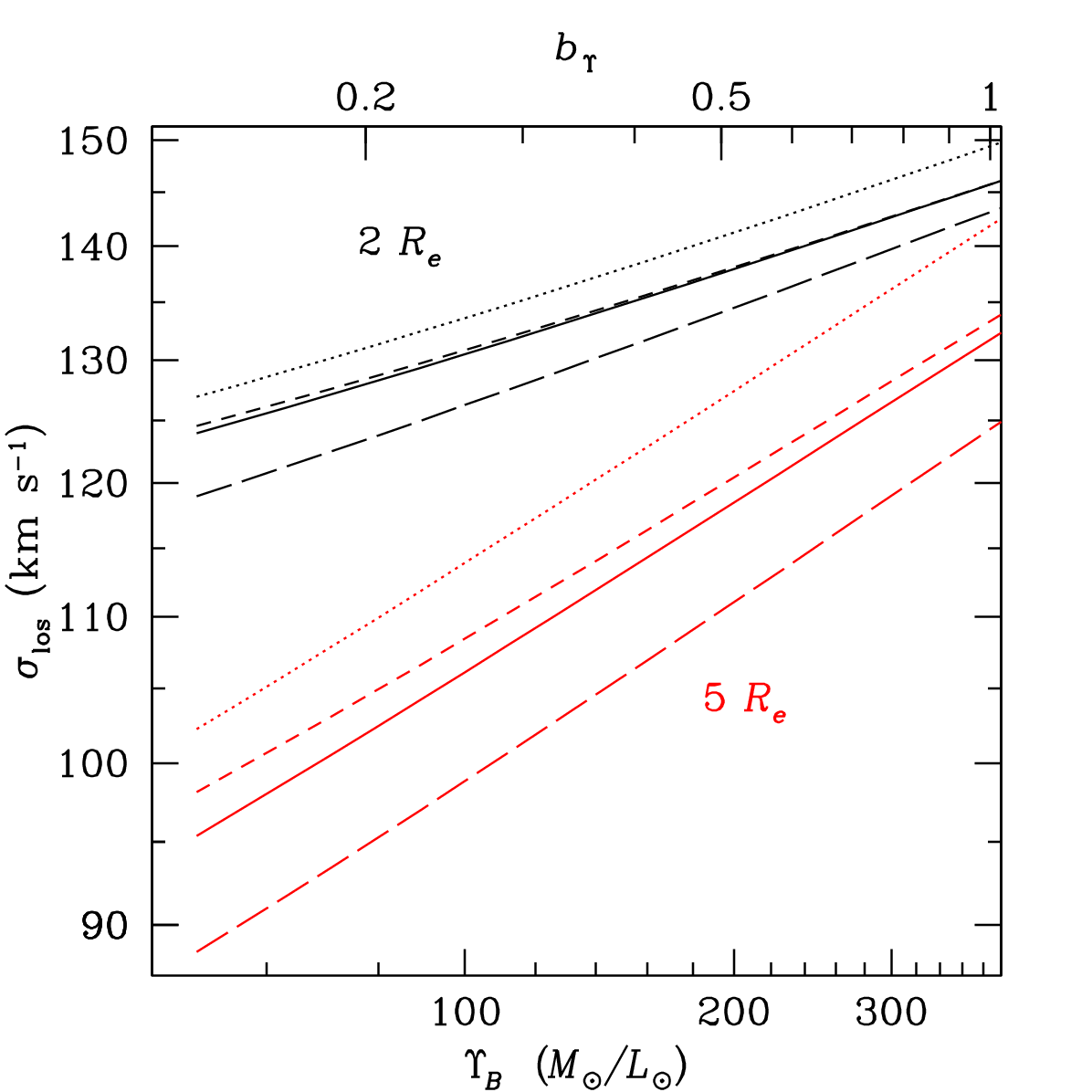}
\caption{Line-of-sight stellar velocity dispersion of our 4-component elliptical
galaxy at 2 (\emph{upper curves}) and 5
(\emph{lower curves}) effective radii, as a function of total mass-to-light
ratio at the virial radius (or its bias relative to the universal value,
upper axis).
The \emph{dotted}, \emph{short-dashed}, \emph{solid} and \emph{long-dashed
curves} represent the
isotropic NFW, isotropic Nav04 and anisotropic (eq.~[\ref{betafit}]) Nav04
models, respectively, the latter two with $r_a/r_v = 0.18$ and 0.018,
respectively.
Same parameters as in Figure~\ref{figcomponents}, in particular no
baryon bias ($b_b=1$), which limits the virial mass-to-light ratios as in
equation~(\ref{gpos}).
}
\label{siglosUps49}
\end{figure}

Reverting now to our standard luminosity $L_B = L_{*,B} = 1.88 \times 
10^{10} {\rm L_\odot}$, we look in
more detail at how the velocity dispersion at a fixed number of effective radii
scales with the total mass-to-light ratio.
Figure~\ref{siglosUps49} shows that the velocity dispersion at 2 and $5\,R_e$
rises slowly with mass-to-light ratio,  
as a power-law of  slope $1/13$
and $1/6$, respectively.
In other words, \emph{the
virial mass-to-light ratio increases very sharply with the measured  
line-of-sight velocity dispersion at a fixed number of effective radii}.
Also, at fixed measured velocity dispersion, the standard
isotropic NFW model produces significantly smaller mass-to-light ratios than
the anisotropic Nav04 model. 
\begin{figure}
\includegraphics[width=\hsize]{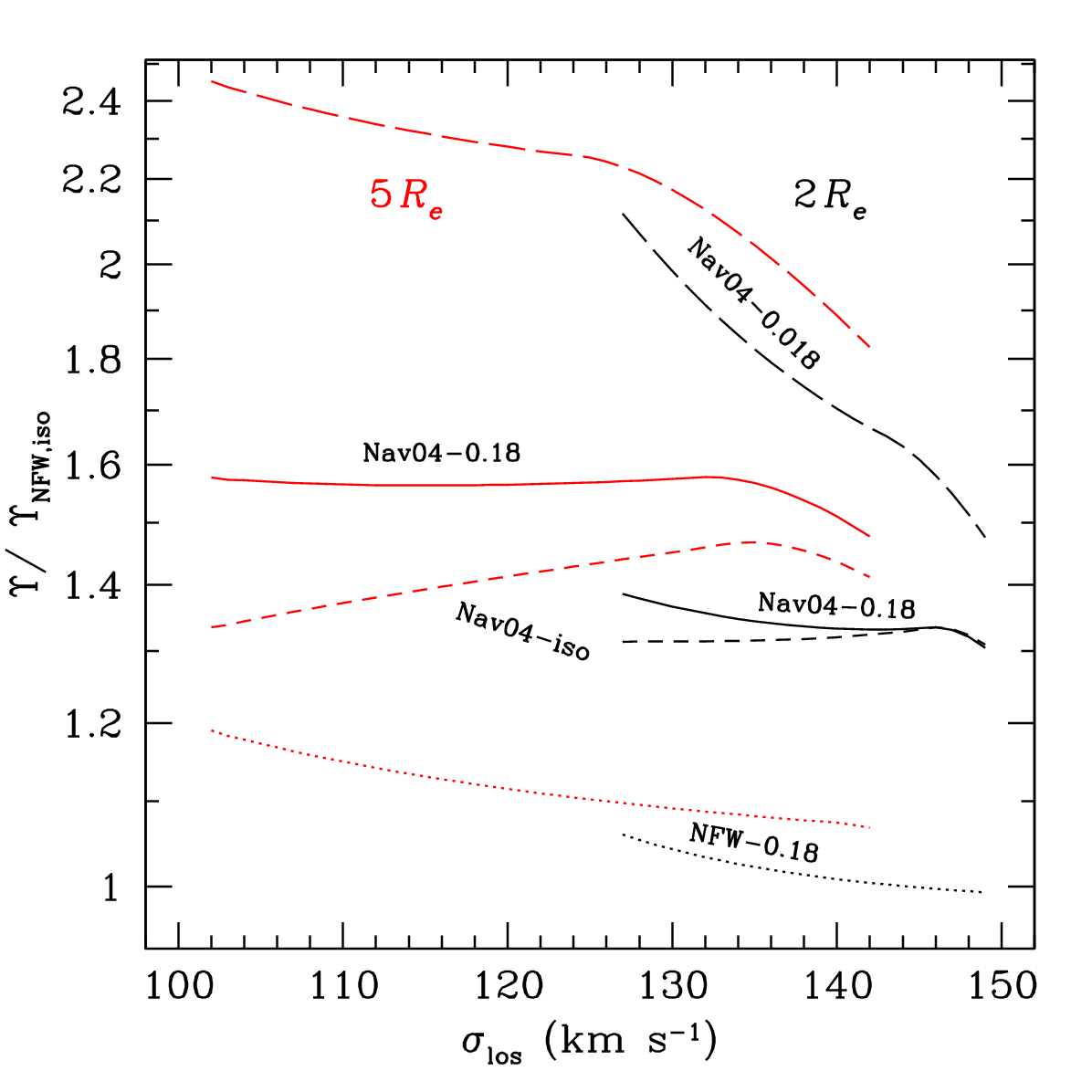}
\caption{Ratio of the mass-to-light ratio inferred from a general model to
the mass-to-light ratio inferred from the isotropic NFW model, as a function of
the measured line-of-sight
velocity dispersion at 5 (\emph{left}) and 2 (\emph{right})
effective radii for a continuous range of virial mass-to-light ratios.
The \emph{short-dashed}, \emph{solid} and \emph{long-dashed
curves} represent the ratios to isotropic NFW mass-to-light ratios for the
isotropic Nav04 and anisotropic (eq.~[\ref{betafit}]) Nav04
models, respectively, the latter two with $r_a/r_v = 0.18$ and 0.018,
respectively, while the \emph{dotted curve} shows the analogous ratio for the
anisotropic NFW model (eq.~[\ref{betafit}] with $r_a = 0.18\,r_v$).
Same parameters as in Figure~\ref{figcomponents}.
}
\label{upsbias} 
\end{figure}

This is quantified in Figure~\ref{upsbias}:
At $5\,R_e$, going from the isotropic NFW model to the isotropic Nav04 model,
the inferred mass-to-light ratio is $\simeq 1.4$ times higher (left
`Nav04-iso' curve), and this factor is roughly independent of the measured
velocity dispersion.  Indeed, as seen in Figure~\ref{ratios}, at $5-20
\,R_e$, 
the velocity
dispersion of the NFW model is a few percent larger than that of the Nav04
model. One therefore 
needs a higher mass-to-light ratio to reproduce a fixed value of
$\sigma_{\rm los}$ with the Nav04 model in comparison with the
less accurate NFW model.
Moreover, going from the isotropic NFW (Nav04) model to the slightly anisotropic
NFW (Nav04) 
model of equation~(\ref{betafit}), the inferred mass-to-light ratio is 1.1
to 1.2 times larger (increasing for smaller measured velocity dispersions),
again as expected since radial anisotropy causes lower values of $\sigma_{\rm
los}$ at large radii (see Fig.~\ref{siglosanis}).  The combined effect of
dark matter model and velocity anisotropy is displayed in the upper
curve of Figure~\ref{upsbias}, and indicates that \emph{line-of-sight
velocity dispersion
measurements lower than $135 \, \rm km \, s^{-1}$ for an $L_*$
elliptical galaxy imply an underestimation of the
mass-to-light 
ratio within the virial radius of a factor of roughly 1.6, 
when using the isotropic NFW model instead of the best slightly anisotropic
models arising from dissipationless 
cosmological $N$-body simulations}.  
Similarly, \emph{the velocity dispersion at $5\,R_e$ produced by the
Nav04 dark matter model with 
the strong radial stellar anisotropy ($r_a = 0.018\,r_v$) seen by
\citeauthor{Dekel+05}
in merger
simulations is the same as that produced by an isotropic NFW dark matter
model with a virial mass-to-light ratio 2.3 times lower}.

Similar patterns are found at
$2\,R_e$ but with considerably smaller effects.
Note that a
positive baryon bias ($b_b>1$) will produce lower velocity dispersions at a
given radius for a given mass-to-light ratio (Fig.~\ref{siglosMbb}), hence the
inferred mass-to-light ratio will be increased relative to the value obtained
for $b_b=1$.

How far out should one measure velocity dispersions?
Table~\ref{uncer} compares the uncertainties at 2 and $5\,R_e$, from
the two effects we discussed above: 1) the observational uncertainty on the
velocity dispersion as inferred from the slopes of the curves in
Figure~\ref{siglosUps49}, and 2) the
uncertainty caused by uncertain total (or 
equivalently dark) mass and
anisotropy profiles, as inferred from Figure~\ref{upsbias}, where we assume
the velocity dispersion profile given in Figure~\ref{siglosUps49} and guess
the measurement errors (see also Sect.~\ref{discus}, below).
\begin{table}
\caption{Relative 
uncertainties on the mass-to-light ratio at the virial radius}
\begin{center}
\tabcolsep 3 pt
\begin{tabular}{lccccc}
\noalign{\hrule}
\\[-7pt]
$R$	& $\sigma_v$ & $\delta \sigma_v$ 
&	$\displaystyle\left({\delta\Upsilon\over \Upsilon}\right)_{\rm obs}$
&	$\displaystyle\left({\delta\Upsilon\over \Upsilon}\right)_{\rm mod}$ 
&	$\displaystyle\left({\delta\Upsilon\over \Upsilon}\right)_{\rm tot}$ \\
&	$(\, \rm km \, s^{-1}$) & $(\, \rm km \, s^{-1})$ & & & \\
\noalign{\hrule}
\\[-9pt]
$2\,R_e$ & 130 & 5 & 0.9 & 1.0 & 1.3 \\
$5\,R_e$ & 106 & 20 & 1.8 & 1.4 & 2.3 \\
$5\,R_e$ & 106 & 10 & 0.7 & 1.4 & 1.6 \\
\noalign{\hrule}
\end{tabular}
\end{center}
\label{uncer}
\end{table}
For assumed measurement errors on $\sigma_{\rm los}$ of $5\,\rm
km \, s^{-1}$ at $2\,R_e$ and $20 \, \rm km \, s^{-1}$ at $5\,R_e$, we end up
with a total factor of 2.3 at $2\,R_e$ and of 3.3 at $5\,R_e$, the latter
factor being reduced to 2.6 if the velocity dispersion at $5\,R_e$ 
can be measured with $10 \, \rm km \, s^{-1}$ accuracy.

Of course, one can do better by combining the measurements of $\sigma_{\rm
los}$ at various radii between 1 and $5\,R_e$.
A quantitative assessment of combining measurements is beyond the scope of the
present paper.
However, inspection of the lower left panel of Figure~\ref{figcomponents},
indicates that \emph{the logarithmic slope of the line-of-sight velocity
dispersion 
profile is roughly independent of the limiting radii, when these lie in
the interval $[1,20]\,R_e$}.
The middle right panel of Figure~\ref{figcomponents} shows that this is not
the case for the (cumulative) mass-to-light ratio: the slope rises with
radius.
In view of the possible $1.3$ (or more) underestimation of the effective radius
of NGC~3379, any error on estimating the effective radius of a galaxy will lead
to biases in analyses based upon the mass-to-light ratio gradient, such as
the study of \cite{Napolitano+05}. 

\begin{figure}
\includegraphics[width=\hsize]{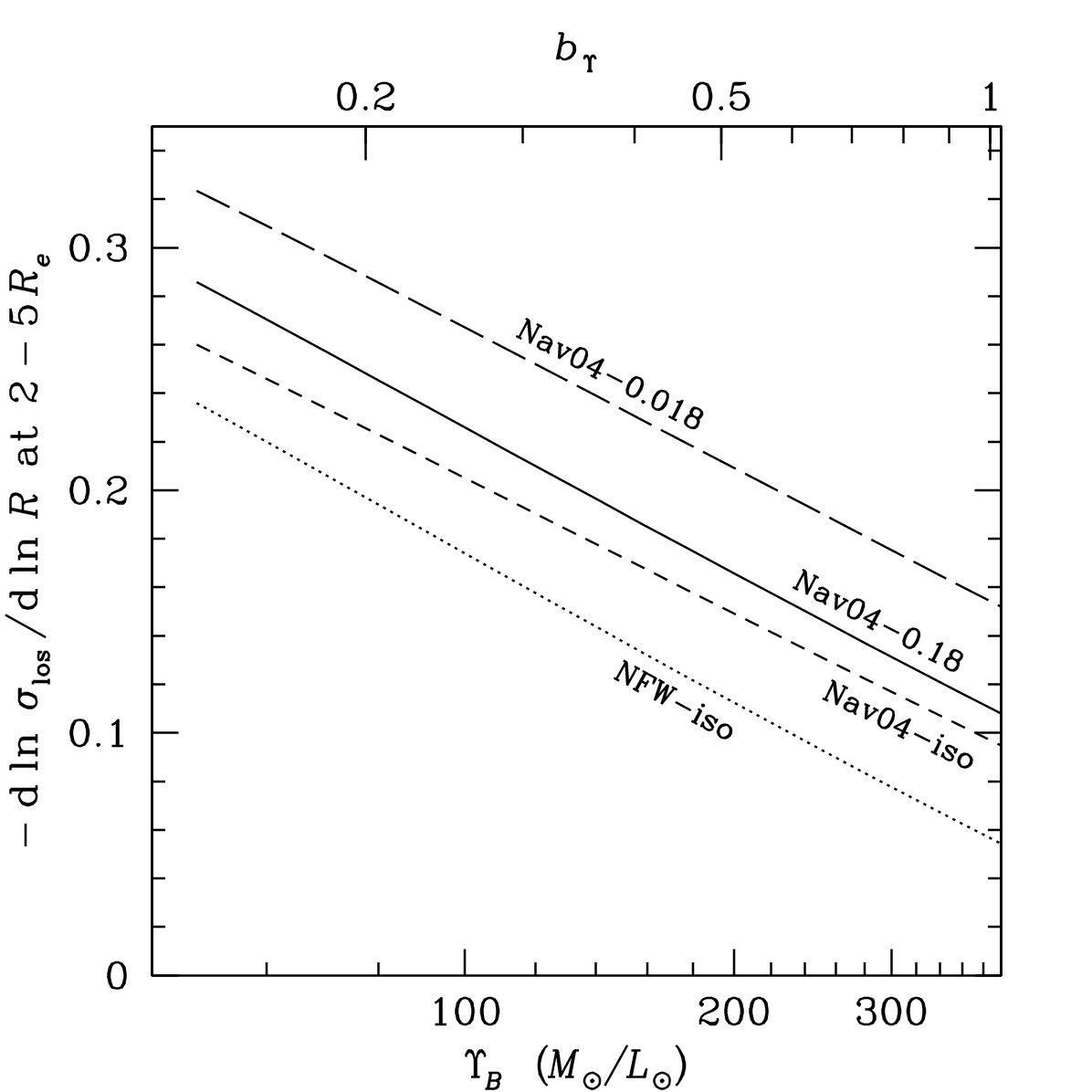}
\caption{Logarithmic slope of the radial profile of the 
line-of-sight stellar velocity dispersion a a function of the total
mass-to-light ratio at
the virial radius.
Same line types as in Fig.~\ref{siglosUps49}.
}
\label{siglosslope} 
\end{figure}

It is much safer to base one's analysis on the
velocity dispersion gradient. 
Figure~\ref{siglosslope} shows that the logarithmic slope of the line-of-sight
velocity dispersion profile decreases linearly with the log of the 
mass-to-light ratio at the virial radius.
Alas, the direct interpretation of this logarithmic slope is complicated by
the uncertainties in the dark mass and stellar anisotropy profiles: for a
given slope of the velocity dispersion profile, these uncertainties lead to a
factor 3 uncertainty on the mass (or mass-to-light ratio) at the virial
radius.
Moreover, as noted earlier (Fig.~\ref{siglosL}) the luminosity of the galaxy
also affects the slope of the line-of-sight velocity dispersion profile.

The large range of mass-to-light ratios at the virial radius that we infer
from \citeauthor{Romanowsky+03}'s 15 orbit solutions (Fig.~\ref{chkRomUps})
suggests that the detailed orbit modelling provides little constraint on the
dark matter content of ellipticals within their virial radii.





\section{Summary and discussion}
\label{discus}
In this paper, a 4-component model of elliptical galaxies has been built to
compare the predictions of cosmological $N$-body simulations with the
observations of elliptical galaxies.
The inner regions are dominated by the stellar component, unless
the inner dark matter density profile is as steep as $r^{-3/2}$, which seems
to disagree with the latest cosmological $N$-body simulations of
\cite{SWTS02}, \cite{Power+03},
\cite{Navarro+04}, \cite{DMS04_rho}, and \cite{Stoehr06}.
Therefore, there is little hope in constraining, through the analysis of
internal kinematics, the inner slope of the dark
matter density profile unless it is as large as $3/2$ in absolute value.

The dark matter component is found to become important at typically 3--5
effective radii. At these radii, the galaxy  surface brightness
is low and one requires the sensitivity of an 8m-class
telescope, and the required precision on velocity dispersion of say $5 \,
\rm km \, s^{-1}$ imposes a decent spectral resolution (with $R=1000$, a
signal-to-noise ratio $S/N > 10$ should lead to $10 \, \rm km \, s^{-1}$
precision on observed
velocity dispersions --- L. Campbell, private communication).
For example, according to the Exposure Time Calculator at the ESO Web
site\footnote{http://www.eso.org/observing/etc},
observations of the giant elliptical galaxy NGC~3379, using the
the S\'ersic surface brightness profile we fit (see Sect.~\ref{UpsRomsec}), 
on the {\sf VLT-UT1 (Antu)} 8m
telescope using the {\sf
FORS2} multi-slit spectrograph at $R=2000$, one can reach $S/N=10$ (per
spectral pixel and per arcsec) in 3 hours at $R = 2\,R_e$.
Among the 16 slits, one can dedicate say 10 for $R = 3\,R_e$ and the combined
spectrum should produce $S/N > 10$ with a precision on velocity dispersion of
order $5 \, \rm km \, s^{-1}$.
%
This measurement at $3\,R_e$ should
constrain the total mass-to-light ratio within the virial radius,  but to
fairly low precision: at $3\,R_e$, one has
${\rm d}\ln\Upsilon_B / {\rm d}\ln \sigma_v \simeq 9$,
yielding an accuracy of a factor of 1.5 only.

Three physical mechanisms are found to decrease the observed line-of-sight
velocity dispersion at 2--5 effective radii: 
\begin{enumerate}
\item
the lower cumulative mass of the
new dark matter models such as that of \cite{Navarro+04} at these radii and
slightly above; 
\item 
radial velocity anisotropy;
\item a possible excess of the
baryon fraction relative to the universal value.
\end{enumerate}

In turn, if we model the observed velocity dispersions at 5 effective radii,
the combined effect of the new dark matter models and velocity
anisotropy is to increase the inferred mass-to-light ratio at the virial
radius by 60\% for slight anisotropy, as observed for the particles in
structures within dissipationless cosmological $N$-body simulations, and by a
factor 2.4 for the strong anisotropy found by \cite{Dekel+05} in simulations
of merging galaxies.

The effect of the velocity anisotropy on the mass at the virial radius is a
direct illustration of the mass / anisotropy degeneracy.
On the other hand, the analysis of the galaxy kinematics at some radius  $R$
requires the knowledge of the mass profile beyond $R$ in theory, but just
a little beyond $R$ in practice, since the mass profile entering the
expression (eq.~[\ref{Isiglosgen}]) for the line of sight velocity dispersion
is weighted by the 
luminosity density profile, which falls relatively fast.
Therefore, the extrapolation of the mass profile out to the virial radius
depends on the details of the mass profile.

In their analysis of halos from dissipationless cosmological $N$-body
 simulations, \cite*{SLM04} showed that the mass profile was recovered to
 high accuracy beyond $0.03\,r_v$.
The difference here, is that the line of
sight velocity dispersion data in ellipticals extends only out to 2 to 5
effective radii, i.e. to less than 6\% of the virial radius, while in
clusters of galaxies, the analogous dispersions extend all the way out to the
virial radius.

One may be tempted to explain the very low ($\Upsilon_B = 33$)  
mass-to-light ratio at the virial
radius reported by \cite{Romanowsky+03} (which, at face value requires the
baryonic fraction within the virial radius to be larger than the universal
value) by this factor of 2.4, which would
raise the mass-to-light ratio at $r_v$ 
to $\Upsilon_B \approx 80$, and to roughly 100 when we use
$r_{102}$ for the virial radius instead of $r_{200}$.
However, this comparison is not fair, because the best fitting orbital
solutions of \citeauthor{Romanowsky+03} are also anisotropic: their orbits
are nearly as radial as the stellar orbits at a few $R_e$ found by
\citeauthor{Dekel+05} (A. Romanowsky, private communication).

Furthermore, the low dispersions of planetary nebulae velocities around
nearby ellipticals measured by \citeauthor{Romanowsky+03}, $\approx 80 \, \rm
km \, s^{-1}$, cannot be reproduced with our models: as seen in
Fig.~\ref{siglosUps49}, we consistently produce velocity dispersions at
$5\,R_e$ above $\sigma_{\rm los}=95 \, \rm km \, s^{-1}$ for
$\Upsilon_B>80$. 
Therefore, resorting to 
radial anisotropy and to the Nav04 dark matter model appear to be
insufficient to explain the low velocity dispersions observed at large radii
by \citeauthor{Romanowsky+03}.
And yet, we found  that some of the orbit solutions
found by these authors lead to high enough mass-to-light ratios
(Fig.~\ref{chkRomUps}), 
as does a  new solution (Romanowsky, private
communication, and right-most pair of points in Fig.~\ref{chkRomUps}).
In fact, the 2nd most massive among the solutions of
\citeauthor{Romanowsky+03} (third pair of points from the right in
Fig.~\ref{chkRomUps}) has $\Upsilon_B(r_{102}) = 70$, and if a Nav04 dark
matter model were used instead of the NFW model, one would get roughly 40\%
larger masses at the virial radius (see Fig.~\ref{upsbias}), i.e. 100, as
roughly expected from cosmological modelling
\citep{MH02,YMvdB03}.
Note that the more massive solutions of \citeauthor{Romanowsky+03} have low
dark matter concentration parameters, compared with the expectations of
\cite{Bullock+01}, revised by \cite{Napolitano+05}, but one solution
($\Upsilon_{B,102}=70$, $c=10$) is consistent with the predictions from
dissipationless cosmological simulations.

\citeauthor{Dekel+05} suggest other ways to bring down the velocity
dispersions: a steeper density profile for the planetary nebulae measured by
\citeauthor{Romanowsky+03}, and viewing triaxial galaxies along their minor
axis.
Another possibility is an incomplete dynamical 
equilibrium of elliptical galaxies, caused by small residual time-variations,
affecting both  the orbit
modelling performed by \citeauthor{Romanowsky+03} and
the Jeans kinematical modelling, as presented here.
Detailed analyses of kinematical observations and of merger simulations are in
progress to help clarify the dark matter content of elliptical galaxies.

















\section*{Acknowledgments}

We thank 
Avishai Dekel,
Daniel Gerbal,
Gastao Lima Neto,
Trevor Ponman,
Jose Maria Rozas,
Elaine Sadler,
and
Felix Stoehr
for useful
conversations and Yi Peng Jing for providing us with the JS--1.5 concentration
parameter from his simulations in digital form.
We also gratefully acknowledge the referee, Aaron Romanowsky, whose
numerous 
insightful comments had a very positive impact on this paper and for
providing unpublished details of his work.
EL\L \
acknowledges hospitality of Institut d'Astrophysique de Paris, where
part of this work was done, while GAM benefited in turn from the hospitality
of the Copernicus Center in Warsaw.
This research was partially supported by the
Polish Ministry of Scientific Research and Information Technology
under grant
1P03D02726
as well as the Jumelage program Astronomie France Pologne of
CNRS/PAN.
This research has benefited from the {\tt Dexter} data extraction applet
developed by the NASA Astrophysics Data System (ADS).

\appendix
\onecolumn

\section{Line-of-sight velocity dispersions for anisotropic models}
\label{appJeans}
\subsection{Radial velocity dispersion}

The general solution to the Jeans equation~(\ref{jeans1}) is
\begin{equation}
\ell (r) \,\sigma_r^2 (r) = {1\over f(r)}\,\int_r^\infty f(s)\, \ell(s)\,
{GM(s)\over s^2} \,{\rm d}s\ ,
\label{jeansgen}
\end{equation}
where $f$ is the solution to
\begin{equation}
{{\rm d}\ln f\over {\rm d}\ln r} = 2\,\beta(r) \ ,
\label{fofbeta}
\end{equation}
i.e.,
\[
f(r) = f(r_1)\,\exp\left [\int_{r_1}^r 2 \,\beta(s)\,{{\rm d}s\over s} \right ]
\]
(see also \citealp{vanderMarel94}).

For isotropic orbits, the Jeans equation~(\ref{jeans1}) trivially leads
($f=1$) to
\begin{equation}
{\ell(r)\,\sigma_r^2 (r)\over G} = \int_r^\infty \ell \,M \,\left
({{\rm d}s\over s^2} \right ) \ .
\label{sigriso}
\end{equation}
For radial orbits, one has \citep{LM03}
\begin{equation}
\ell(r)\,\sigma_r^2 (r) = {G\over r^2}\,\int_r^\infty
\ell\,M\, {\rm d}s \ .
\label{sigrradial}
\end{equation}
%
%
%
%
Also, with the Osipkov-Merritt anisotropy (eq.~[\ref{betaOM}]) one gets with
$f = r^2+r_a^2$
\begin{equation}
\ell(r)\,\sigma_r^2 (r) = {G\over r^2 +r_a^2}\,\int_r^\infty
\ell\,M\,\left ({s^2+r_a^2 \over s^2}\right)\,{\rm d}s\ .
\label{sigrOM}
\end{equation}
Similarly, with the anisotropy profile of equation~(\ref{betafit}), shown as
the solid curve in Figure~\ref{betasims},
one gets with equations~(\ref{jeansgen}) and (\ref{fofbeta}) and $f=r+r_a$
\begin{equation}
\ell(r)\,\sigma_r^2 (r) = {G\over r +r_a}\,\int_r^\infty
\ell\,M\,\left ({s+r_a \over s^2}\right)\,{\rm d}s\ .
\label{sigrnewbeta}
\end{equation}

\subsection{Line-of-sight velocity dispersion}

\subsubsection{General formulas}

Projecting the velocity ellipsoid along the line-of-sight, one finds that
 the line-of-sight velocity
dispersion is \citep{BM82}
\begin{equation}
{1\over 2} I(R)\,\sigma_{\rm los}^2 (R) = \int_R^\infty {\ell\sigma_r^2
\,r\,{\rm d}r\over \sqrt{r^2\!-\!R^2}} - R^2 \int_R^\infty
{\beta\,\ell\sigma_r^2\,{\rm d}r\over r\,\sqrt{r^2\!-\!R^2}}
\ .
\label{siglos1}
\end{equation}


Inserting equations~(\ref{jeansgen}) and
(\ref{fofbeta}) into equation~(\ref{siglos1}) and
inverting the order of integration, one obtains the general expression
\begin{equation}
{I(R)\,\sigma_{\rm los}^2 (R)\over G}
=
2\,\int_R^\infty {f\,\ell\,M\over s^2}\,{\rm d}s\int_R^s {1\over f}\,{r\,{\rm d}r\over
\sqrt{r^2-R^2}} - R^2\,
\int_R^\infty {f\,\ell\,M\over s^2}\,{\rm d}s\int_R^s {{\rm d}f/{\rm d}r\over f^2}\,{{\rm d}r\over
\sqrt{r^2-R^2}} \ .
\label{siglosgenf}
\end{equation}
The inner integrals of equation~(\ref{siglosgenf}) are closed for only
several realistic\footnote{we reject $f=\sqrt{r^2-a^2}$ and
$f=1/\sqrt{r^2-a^2}$, as both lead to unphysical $\beta>1$ near $r=a$.}
forms of $f$: 
\begin{enumerate}
\item $f = r^{2\,\beta}, which implies \beta =
\hbox{cst}$; 
\item $f = (r+a)^\alpha$, with $\alpha = 1$ or 2, which implies
$\beta = (\alpha/2)\,r/(r+a)$; 
\item $f=(r^2+a^2)^\alpha$, again with $\alpha =
1$ or 2, which implies $\beta = \alpha\,r^2/(r^2+a^2)$ --- hence, $\alpha=1$
corresponds to Osipkov-Merritt \citep{Osipkov79,Merritt85_df} anisotropy.
\end{enumerate}

For the anisotropy profile of equation~(\ref{betafit}), one obtains from
equation~(\ref{siglosgenf}), for $R \neq r_a$,
after inverting the order of integration, 
with plenty of algebra
\begin{eqnarray}
{I(R)\,\sigma_{\rm los}^2 (R)\over G} 
&=& 2\,\int_R^\infty \left ({r+r_a\over r^2} \right )
\hbox{cosh}^{-1} \left ({r\over R} \right )\,\ell (r) \,M(r)\,{\rm d}r 
+ {R^2\over r_a^2-R^2}\,\int_R^\infty {\sqrt{r^2-R^2}\over r^2}\,
\ell (r) \,M(r)\,{\rm d}r \nonumber \\
&\mbox{}& \quad 
 -\hbox{sgn}\left ({r_a\over R}-1\right)\, r_a\,{2\,r_a^2-R^2 \over \left
 |r_a^2-R^2\right|^{3/2}}\, 
\int_R^\infty \left ({r+r_a\over r^2}\right)\,
\hbox{C}^{-1} \left ({r_a r/R + R\over r+r_a} \right )
\ell (r) \,M(r)\,{\rm d}r \ ,
\label{siglosnewbeta}
\end{eqnarray}
where $\hbox{C}^{-1} (x) = \cosh^{-1} x$ for $r_a > R$
and $\hbox{C}^{-1} (x) = \cos^{-1} x$ for $r_a < R$.\footnote{Here and below,
we use the positive definitions of $\cos^{-1} x$ and $\cosh^{-1} x$.}
For $R=r_a$, we similarly obtain
\begin{equation}
{I(r_a)\,\sigma_{\rm los}^2 (r_a)\over G} 
=2\,\int_{r_a}^\infty 
\left ({r+r_a\over r^2}\right )\,
\hbox{cosh}^{-1} \left ({r\over r_a} \right )
\,\ell (r)\,M(r)\,{\rm d}r  
-{1\over3}\,\int_{r_a}^\infty \,
\left ({8\,r_a+7\,r\over r^2}\right )
\,\sqrt{r-r_a\over r+r_a}\,\ell(r) M(r)\,{\rm d}r \ .
\end{equation}

In the limit of isotropy ($r_a \to \infty$), we recover (after some algebra
on eq.~[\ref{siglosnewbeta}])
the standard 
line-of-sight velocity dispersion for isotropic systems (Prugniel \&
Simien 1997):
\begin{equation}
I(R) \,\sigma_{\rm los}^2 (R) = 2\,G\,\int_R^\infty 
{\sqrt{r^2\!-\!R^2}\over r^2}\ \ell(r) \,M(r)\,{\rm d}r \ ,
\label{siglosiso}
\end{equation}
which can also be easily obtained through equations~(\ref{sigriso}) and
(\ref{siglos1}), again after inversion of the order of integration.

For Osipkov-Merritt anisotropy (eq.~[\ref{betaOM}]) one finds, in a similar
fashion,
\begin{equation}
{I(R)\,\sigma_{\rm los}^2 (R)\over G} 
= {2 \,r_a^2 + R^2 \over (r_a^2+R^2)^{3/2}}
\int_R^\infty \left ({r^2+r_a^2 \over r^2}\right )
\tan^{-1} \sqrt{r^2\!-\!R^2\over r_a^2\!+\!R^2}\ \ell (r) \, M(r) \,{\rm d}r
- \,{R^2 \over r_a^2\!+\!R^2}
\int_R^\infty {\sqrt{r^2\!-\!R^2}  \over r^2 } \ \ell (r) \, M(r) \,{\rm d}r \ .
\label{siglosOM}
\end{equation}

For constant anisotropy orbits, the same procedure yields
\begin{eqnarray}
{I(R)\,\sigma_{\rm los}^2 (R)\over G}  &=&
R^{1-2\beta}\,
\left [\left ({3\over2}-\beta \right )\sqrt{\pi}\,
{\Gamma(\beta-1/2)\over \Gamma(\beta)}
\int_R^\infty \ell(r) \,M(r)\,{\rm d}r
+ \beta \int_R^\infty B(R^2/r^2,\beta+1/2,1/2)\,\ell(r) \,M(r)\,{\rm d}r
\right. \nonumber \\
&\mbox{}&\quad \left.
- \int_R^\infty
 B(R^2/r^2,\beta-1/2,1/2)\,\ell(r) \,M(r)\,{\rm d}r \right] \ ,
\end{eqnarray}
where $B(x,a,b) = \int_0^x t^{a-1}(1-t)^{b-1}{\rm d}t$ 
is the incomplete beta function.
For purely radial orbits, one can obtain the simpler formula
\begin{equation}
{I(R)\,\sigma_{\rm los}^2 (R)\over G}  =
{\pi\over 2\,R}\,\int_R^\infty \ell(r) \,M(r)\,{\rm d}r
- \int_R^\infty {\sqrt{r^2-R^2}\over r^2}\ \ell(r) \,M(r)\,{\rm d}r
-{1\over R}\,\int_R^\infty \sin^{-1} \left ({R\over r}\right )\,
\ell(r) \,M(r)\,{\rm d}r \ .
\label{siglosradial}
\end{equation}

Therefore, for our choice of simple anisotropy profiles, we can generally
write
\begin{equation}
I(R)\,\sigma_{\rm los}^2 (R) = 2\,G\,
\int_R^\infty K \left ({r\over R}, {r_a\over R} \right
)\,\ell(r)\,M(r)\,{dr\over r} \ ,
\label{Isiglosgen}
\end{equation}
where the kernel $K$ can be expressed for our five anisotropy models as
\begin{equation}
K(u,u_a) = \left \{
\begin{array}{ll}
\displaystyle
{1\over2}\,u^{2\beta-1}\left [
\left ({3\over2}-\beta \right )\sqrt{\pi}\,
{\Gamma(\beta-1/2)\over \Gamma(\beta)}
+\beta B \left ({1\over u^2},\beta+{1\over 2},{1\over 2} \right)
- B\left({1\over u^2},\beta-{1\over 2},{1\over 2}\right) 
\right ]
& \hbox{(cst-$\beta$)} \ , \\
\\
\displaystyle
\sqrt{1-{1\over u^2}}	& \hbox{(isotropic)} \ , \\
\\
\displaystyle
{\pi\over 4}\,u
- {1\over2}\,\sqrt{1-{1 \over u^2}}
- {u\over 2}\,\sin^{-1}\left ({1\over u}\right )	& \hbox{(radial)} \ ,
\\
\\
\displaystyle
{u_a^2+1/2\over  \left (u_a^2+1\right)^{3/2}}\,\left ({u^2+u_a^2\over
u}\right )\,\tan^{-1} \sqrt{u^2-1\over u_a^2+1} \,-\,{1/2\over
u_a^2+1}\,\sqrt{1-1/ u^2}
& \hbox{(Osipkov-Merritt)} \ , \\
\\
\displaystyle
{1/2 \over u_a^2-1}\,\sqrt{1-{1\over u^2}}
+ 
\left ({1+{u_a \over u}}\right) \cosh^{-1} u \\
\displaystyle 
\mbox{} \quad - \hbox{sgn}\left (u_a-1\right)
u_a {u_a^2-1/2\over \left (u_a^2-1\right )^{3/2}}
\left (1+{u_a\over u} \right )
\hbox{C}^{-1} \left ({u_a u + 1\over u+u_a} \right )
& (\hbox{eq.~[\ref{betafit}]}, u_a \neq 1) \ ,\\
\\
\displaystyle
\left ({1+{1 \over u}}\right)\,\cosh^{-1} u
-{1\over6}\,\left ({8\over u}+7\right)\,\sqrt{u-1\over u+1}
& (\hbox{eq.~[\ref{betafit}]}, u_a = 1) \ ,\\
\end{array}
\right.
\label{sigallker}
\end{equation}
where 
$B$ is the incomplete beta function,
 $\hbox{C}^{-1} (x) = \cosh^{-1} x$ for $u_a > 1$
and $\hbox{C}^{-1} (x) = \cos^{-1} x $ for $u_a < 1$.

\textcolor{red}{Next paragraph from Addendum \citep{ML06b}}: 
In practice, if a programming language does not provide the incomplete Beta
function, but only the regularized incomplete Beta function, $I(x,a,b) =
B(x,a,b)/B(a,b)$, then the top line of equation~(A16) (cst-$\beta$) becomes: 
\[
K(u) = 
{\sqrt{1-1/u^2} \over 1-2\,\beta} + {\sqrt{\pi}\over 2}\,{\Gamma(\beta-1/2)
  \over \Gamma(\beta)} \,\left ({3\over2}-\beta\right)\,u^{2\beta-1}\,\left
[1-I\left ({1\over u^2},\beta+{1\over2},{1\over2}\right)\right] \ . 
\]
For $\beta < -1/2$, one can use the recursion relation (e.g., 
\citealp{ZS65},
eq. [26.5.16]) 
\[
I(x,a,b) = { \Gamma[a+b] \over \Gamma(a+1)\,\Gamma(b) }\,x^a\,(1-x)^b + I(x,a+1,b) \ .
\]

\subsubsection{Application to cold dark matter profiles
and a S\'ersic luminosity profile}

The total mass at a given radius is the sum of masses from the dark, stellar,
gas and black hole components:
\begin{eqnarray}
M(r) &=& M_t\,\widetilde M(r/a_S) \label{Mtotofr}\\
\widetilde M(x) &=&
f_d \,\widetilde M_d(\eta x) + f_* \,{\widetilde L_3(x) \over
\widetilde L_3(x_v)} + f_g \,{\widetilde M_g(\xi x) \over \widetilde
M_g(\xi x_v)} + g_{\rm BH} \,f_* 
\ ,
\label{Mtottilde}
\end{eqnarray}
where we used equations (\ref{L3ofr}), (\ref{fracBH}),
(\ref{gofc}), (\ref{Mdarkofr}), (\ref{mtildedark}),
(\ref{mgas2}), and (\ref{gastotot}), as well as the
component fractions at the virial radius, $f_i$, from equations
(\ref{fstars})-(\ref{fgas}),
where $x = r/a_S$, $x_v = r_v/a_S$, 
$\eta = a_S / a_d$, $\xi = a_S/r_c = q/b^m$,
the deprojected luminosity $\widetilde L_3$ of the S\'ersic profile is given in
equation~(\ref{L3tilde}), and the dimensionless gas profile $\widetilde M_g$
is given in equation~(\ref{mtildegasapx}) for $r \leq r_v$ or
(\ref{mtildegas2}) for $r > r_v$. 
One can set $f_b = f_* = \Upsilon_* = 0$ if one wishes to study the case of
an NFW, JS--1.5, or Nav04
\emph{potential}, i.e. assuming that the total matter density profile is the 
NFW, JS--1.5, or Nav04 density profile.

Using equations~(\ref{Isersic}),
(\ref{nuofr}), 
(\ref{nuofx}), 
(\ref{nu1}),
(\ref{Isiglosgen}),
(\ref{Mtotofr}) and (\ref{Mtottilde})
the line-of-sight velocity dispersion of component $i$ can be written
\begin{equation}
\sigma_{{\rm los},i}^2 (R) = {\Gamma(2m)\over \Gamma[(3\!-\!p)m]}
\,{r_v \over a_S} \, V_v^2\,\exp\left (X^{1/m} \right ) 
\,\int_X^\infty K \left ({x \over X},{x_a \over X} \right)
\,\widetilde \ell(x) \, \widetilde M_i(x) \, {{\rm
d}x \over x} \ ,
\label{sigall}
\end{equation}
where $X = R/a_S$, $x_a=r_a/a_S$, and the kernel $K$ is given in
equation~(\ref{sigallker}).
Then $\sigma_{\rm los}^2(R) = \sum_i \sigma_{{\rm los},i}^2 (R)$.

\twocolumn

\bibliography{master}

\begin{thebibliography}{91}
\expandafter\ifx\csname natexlab\endcsname\relax\def\natexlab#1{#1}\fi

\bibitem[{{Bertin} {et~al.}(2002){Bertin}, {Ciotti}, \& {Del
  Principe}}]{BCDP02}
{Bertin} G., {Ciotti} L., {Del Principe} M., 2002, \aap, 386, 149

\bibitem[{{Binney} \& {Mamon}(1982)}]{BM82}
{Binney} J., {Mamon} G.~A., 1982, \mnras, 200, 361

\bibitem[{{Blanton} {et~al.}(2003){Blanton}, {Hogg}, {Bahcall}, {Brinkmann},
  {Britton}, {Connolly}, {Csabai}, {Fukugita}, {Loveday}, {Meiksin}, {Munn},
  {Nichol}, {Okamura}, {Quinn}, {Schneider}, {Shimasaku}, {Strauss}, {Tegmark},
  {Vogeley}, \& {Weinberg}}]{Blanton+03}
{Blanton} M.~R., {Hogg} D.~W., {Bahcall} N.~A., {Brinkmann} J., {Britton} M.,
  {Connolly} A.~J., {Csabai} I., {Fukugita} M., {Loveday} J., {Meiksin} A.,
  {Munn} J.~A., {Nichol} R.~C., {Okamura} S., {Quinn} T., {Schneider} D.~P.,
  {Shimasaku} K., {Strauss} M.~A., {Tegmark} M., {Vogeley} M.~S., {Weinberg}
  D.~H., 2003, \apj, 592, 819

\bibitem[{{Borriello} {et~al.}(2003){Borriello}, {Salucci}, \&
  {Danese}}]{BSD03}
{Borriello} A., {Salucci} P., {Danese} L., 2003, \mnras, 341, 1109

\bibitem[{{Brown} \& {Bregman}(2001)}]{BB01}
{Brown} B.~A., {Bregman} J.~N., 2001, \apj, 547, 154

\bibitem[{{Bullock} {et~al.}(2001){Bullock}, {Kolatt}, {Sigad}, {Somerville},
  {Kravtsov}, {Klypin}, {Primack}, \& {Dekel}}]{Bullock+01}
{Bullock} J.~S., {Kolatt} T.~S., {Sigad} Y., {Somerville} R.~S., {Kravtsov}
  A.~V., {Klypin} A.~A., {Primack} J.~R., {Dekel} A., 2001, \mnras, 321, 559

\bibitem[{{Caon} {et~al.}(1993){Caon}, {Capaccioli}, \& {D'Onofrio}}]{CCDO93}
{Caon} N., {Capaccioli} M., {D'Onofrio} M., 1993, \mnras, 265, 1013

\bibitem[{{Capaccioli} {et~al.}(1990){Capaccioli}, {Held}, {Lorenz}, \&
  {Vietri}}]{CHLV90}
{Capaccioli} M., {Held} E.~V., {Lorenz} H., {Vietri} M., 1990, \aj, 99, 1813

\bibitem[{{Carlberg} {et~al.}(1997){Carlberg}, {Yee}, {Ellingson}, {Morris},
  {Abraham}, {Gravel}, {Pritchet}, {Smecker-Hane}, {Hartwick}, {Hesser},
  {Hutchings}, \& {Oke}}]{Carlberg+97_mofr}
{Carlberg} R.~G., {Yee} H. K.~C., {Ellingson} E., {Morris} S.~L., {Abraham} R.,
  {Gravel} P., {Pritchet} C.~J., {Smecker-Hane} T., {Hartwick} F. D.~A.,
  {Hesser} J.~E., {Hutchings} J.~B., {Oke} J.~B., 1997, \apjl, 485, L13

\bibitem[{{Col{\'{\i}}n} {et~al.}(2000){Col{\'{\i}}n}, {Klypin}, \&
  {Kravtsov}}]{CKK00}
{Col{\'{\i}}n} P., {Klypin} A.~A., {Kravtsov} A.~V., 2000, \apj, 539, 561

\bibitem[{{Col{\'{\i}}n} {et~al.}(1999){Col{\'{\i}}n}, {Klypin}, {Kravtsov}, \&
  {Khokhlov}}]{CKKK99}
{Col{\'{\i}}n} P., {Klypin} A.~A., {Kravtsov} A.~V., {Khokhlov} A.~M., 1999,
  \apj, 523, 32

\bibitem[{{Colina} {et~al.}(1996){Colina}, {Bohlin}, \& {Castelli}}]{CBC96}
{Colina} L., {Bohlin} R.~C., {Castelli} F., 1996, \aj, 112, 307

\bibitem[{{Cox} {et~al.}(2005){Cox}, {Jonsson}, {Primack}, \&
  {Somerville}}]{CJPS05}
{Cox} T.~J., {Jonsson} P., {Primack} J.~R., {Somerville} R.~S., 2005, \mnras,
  submitted, arXiv:astro-ph/0503201

\bibitem[{{de Vaucouleurs}(1948)}]{deVaucouleurs48}
{de Vaucouleurs} G., 1948, Annales d'Astrophysique, 11, 247

\bibitem[{{de Vaucouleurs} \& {Capaccioli}(1979)}]{dVC79}
{de Vaucouleurs} G., {Capaccioli} M., 1979, \apjs, 40, 699

\bibitem[{{de Vaucouleurs} {et~al.}(1991){de Vaucouleurs}, {de Vaucouleurs},
  {Corwin}, {Buta}, {Paturel}, \& {Fouqu\'e}}]{RC3}
{de Vaucouleurs} G., {de Vaucouleurs} A., {Corwin} J.~R., {Buta} R.~J.,
  {Paturel} G., {Fouqu\'e} P., 1991, Third Reference Catalogue of Bright
  Galaxies. Springer, New York

\bibitem[{{Dekel} {et~al.}(2005){Dekel}, {Stoehr}, {Mamon}, {Cox}, {Novak}, \&
  {Primack}}]{Dekel+05}
{Dekel} A., {Stoehr} F., {Mamon} G.~A., {Cox} T.~J., {Novak} G.~S., {Primack}
  J.~R., 2005, \nat, 437, 707, arXiv:astro-ph/0501622

\bibitem[{{Diaferio}(1999)}]{Diaferio99}
{Diaferio} A., 1999, \mnras, 309, 610

\bibitem[{{Diemand} {et~al.}(2004{\natexlab{a}}){Diemand}, {Moore}, \&
  {Stadel}}]{DMS04_rho}
{Diemand} J., {Moore} B., {Stadel} J., 2004{\natexlab{a}}, \mnras, 353, 624

\bibitem[{{Diemand} {et~al.}(2004{\natexlab{b}}){Diemand}, {Moore}, \&
  {Stadel}}]{DMS04_vel}
---, 2004{\natexlab{b}}, \mnras, 352, 535

\bibitem[{{Eke} {et~al.}(1996){Eke}, {Cole}, \& {Frenk}}]{ECF96}
{Eke} V.~R., {Cole} S., {Frenk} C.~S., 1996, \mnras, 282, 263

\bibitem[{{Faber} \& {Jackson}(1976)}]{FJ76}
{Faber} S.~M., {Jackson} R.~E., 1976, \apj, 204, 668

\bibitem[{{Faber} {et~al.}(1997){Faber}, {Tremaine}, {Ajhar}, {Byun},
  {Dressler}, {Gebhardt}, {Grillmair}, {Kormendy}, {Lauer}, \&
  {Richstone}}]{Faber+97}
{Faber} S.~M., {Tremaine} S., {Ajhar} E.~A., {Byun} Y., {Dressler} A.,
  {Gebhardt} K., {Grillmair} C., {Kormendy} J., {Lauer} T.~R., {Richstone} D.,
  1997, \aj, 114, 1771

\bibitem[{{Fukushige} \& {Makino}(1997)}]{FM97}
{Fukushige} T., {Makino} J., 1997, \apjl, 477, L9

\bibitem[{{Gerhard} {et~al.}(1998){Gerhard}, {Jeske}, {Saglia}, \&
  {Bender}}]{GJSB98}
{Gerhard} O., {Jeske} G., {Saglia} R.~P., {Bender} R., 1998, \mnras, 295, 197

\bibitem[{{Gerhard} {et~al.}(2001){Gerhard}, {Kronawitter}, {Saglia}, \&
  {Bender}}]{GKSB01}
{Gerhard} O., {Kronawitter} A., {Saglia} R.~P., {Bender} R., 2001, \aj, 121,
  1936

\bibitem[{{Gerhard}(1993)}]{Gerhard93}
{Gerhard} O.~E., 1993, \mnras, 265, 213

\bibitem[{{Ghigna} {et~al.}(2000){Ghigna}, {Moore}, {Governato}, {Lake},
  {Quinn}, \& {Stadel}}]{Ghigna+00}
{Ghigna} S., {Moore} B., {Governato} F., {Lake} G., {Quinn} T., {Stadel} J.,
  2000, \apj, 544, 616

\bibitem[{{Gnedin} {et~al.}(2004){Gnedin}, {Kravtsov}, {Klypin}, \&
  {Nagai}}]{GKKN04}
{Gnedin} O.~Y., {Kravtsov} A.~V., {Klypin} A.~A., {Nagai} D., 2004, \apj, 616,
  16

\bibitem[{{Goudfrooij} {et~al.}(1994){Goudfrooij}, {Hansen}, {Jorgensen},
  {Norgaard-Nielsen}, {de Jong}, \& {van den Hoek}}]{Goudfrooij+94}
{Goudfrooij} P., {Hansen} L., {Jorgensen} H.~E., {Norgaard-Nielsen} H.~U., {de
  Jong} T., {van den Hoek} L.~B., 1994, \aaps, 104, 179

\bibitem[{{Graham} \& {Driver}(2005)}]{GD05}
{Graham} A.~W., {Driver} S.~P., 2005, Publications of the Astronomical Society
  of Australia, 22, 118

\bibitem[{{Graham} \& {Guzm{\' a}n}(2003)}]{GG03}
{Graham} A.~W., {Guzm{\' a}n} R., 2003, \aj, 125, 2936

\bibitem[{{Guzik} \& {Seljak}(2002)}]{GS02}
{Guzik} J., {Seljak} U., 2002, \mnras, 335, 311

\bibitem[{{H{\" a}ring} \& {Rix}(2004)}]{HR04}
{H{\" a}ring} N., {Rix} H., 2004, \apjl, 604, L89

\bibitem[{{Hernquist}(1990)}]{Hernquist90}
{Hernquist} L., 1990, \apj, 356, 359

\bibitem[{{Hoekstra} {et~al.}(2004){Hoekstra}, {Yee}, \& {Gladders}}]{HYG04}
{Hoekstra} H., {Yee} H.~K.~C., {Gladders} M.~D., 2004, \apj, 606, 67

\bibitem[{{Illingworth}(1977)}]{Illingworth77}
{Illingworth} G., 1977, \apjl, 218, L43

\bibitem[{{Jing} \& {Suto}(2000)}]{JS00}
{Jing} Y.~P., {Suto} Y., 2000, \apjl, 529, L69

\bibitem[{{Jorgensen} {et~al.}(1996){Jorgensen}, {Franx}, \&
  {Kjaergaard}}]{JFK96}
{Jorgensen} I., {Franx} M., {Kjaergaard} P., 1996, \mnras, 280, 167

\bibitem[{{Katgert} {et~al.}(2004){Katgert}, {Biviano}, \& {Mazure}}]{KBM04}
{Katgert} P., {Biviano} A., {Mazure} A., 2004, \apj, 600, 657

\bibitem[{{Kormendy} {et~al.}(1997){Kormendy}, {Bender}, {Magorrian},
  {Tremaine}, {Gebhardt}, {Richstone}, {Dressler}, {Faber}, {Grillmair}, \&
  {Lauer}}]{Kormendy+97}
{Kormendy} J., {Bender} R., {Magorrian} J., {Tremaine} S., {Gebhardt} K.,
  {Richstone} D., {Dressler} A., {Faber} S.~M., {Grillmair} C., {Lauer} T.~R.,
  1997, \apjl, 482, L139

\bibitem[{{Kronawitter} {et~al.}(2000){Kronawitter}, {Saglia}, {Gerhard}, \&
  {Bender}}]{KSGB00}
{Kronawitter} A., {Saglia} R.~P., {Gerhard} O., {Bender} R., 2000, \aaps, 144,
  53

\bibitem[{{Lima Neto} {et~al.}(1999){Lima Neto}, {Gerbal}, \& {M{\'
  a}rquez}}]{LGM99}
{Lima Neto} G.~B., {Gerbal} D., {M{\' a}rquez} I., 1999, \mnras, 309, 481

\bibitem[{{Liske} {et~al.}(2003){Liske}, {Lemon}, {Driver}, {Cross}, \&
  {Couch}}]{Liske+03}
{Liske} J., {Lemon} D.~J., {Driver} S.~P., {Cross} N.~J.~G., {Couch} W.~J.,
  2003, \mnras, 344, 307

\bibitem[{{Loewenstein} \& {White}(1999)}]{LW99}
{Loewenstein} M., {White} R.~E., 1999, \apj, 518, 50

\bibitem[{{{\L}okas} \& {Hoffman}(2001)}]{LH01}
{{\L}okas} E., {Hoffman} Y., 2001, in The Identification of Dark Matter,
  {Spooner} N. J.~C., {Kudryavtsev} V., eds., World Scientific, Singapore, pp.
  121--126, arXiv:astro-ph/0011295

\bibitem[{{{\L}okas} \& {Mamon}(2001)}]{LM01}
{{\L}okas} E.~L., {Mamon} G.~A., 2001, \mnras, 321, 155

\bibitem[{{{\L}okas} \& {Mamon}(2003)}]{LM03}
---, 2003, \mnras, 343, 401

\bibitem[{{M{\' e}ndez} {et~al.}(2001){M{\' e}ndez}, {Riffeser}, {Kudritzki},
  {Matthias}, {Freeman}, {Arnaboldi}, {Capaccioli}, \& {Gerhard}}]{Mendez+01}
{M{\' e}ndez} R.~H., {Riffeser} A., {Kudritzki} R.-P., {Matthias} M., {Freeman}
  K.~C., {Arnaboldi} M., {Capaccioli} M., {Gerhard} O.~E., 2001, \apj, 563, 135

\bibitem[{{Magorrian} {et~al.}(1998){Magorrian}, {Tremaine}, {Richstone},
  {Bender}, {Bower}, {Dressler}, {Faber}, {Gebhardt}, {Green}, {Grillmair},
  {Kormendy}, \& {Lauer}}]{Magorrian+98}
{Magorrian} J., {Tremaine} S., {Richstone} D., {Bender} R., {Bower} G.,
  {Dressler} A., {Faber} S.~M., {Gebhardt} K., {Green} R., {Grillmair} C.,
  {Kormendy} J., {Lauer} T., 1998, \aj, 115, 2285

\bibitem[{{Mamon} \& {{\L}okas}(2005)}]{ML05a}
{Mamon} G.~A., {{\L}okas} E.~L., 2005, \mnras, 362, 95

\bibitem[{{Mamon} \& {{\L}okas}(2006)}]{ML06b}
---, 2006, \mnras, 370, 1582

\bibitem[{{Marinoni} \& {Hudson}(2002)}]{MH02}
{Marinoni} C., {Hudson} M.~J., 2002, \apj, 569, 101

\bibitem[{{Merritt}(1985)}]{Merritt85_df}
{Merritt} D., 1985, \mnras, 214, 25P

\bibitem[{{Merritt}(1987)}]{Merritt87}
---, 1987, \apj, 313, 121

\bibitem[{{Moore} {et~al.}(1999){Moore}, {Quinn}, {Governato}, {Stadel}, \&
  {Lake}}]{Moore+99}
{Moore} B., {Quinn} T., {Governato} F., {Stadel} J., {Lake} G., 1999, \mnras,
  310, 1147

\bibitem[{{Napolitano} {et~al.}(2005){Napolitano}, {Capaccioli}, {Romanowsky},
  {Douglas}, {Merrifield}, {Kuijken}, {Arnaboldi}, {Gerhard}, \&
  {Freeman}}]{Napolitano+05}
{Napolitano} N.~R., {Capaccioli} M., {Romanowsky} A.~J., {Douglas} N.~G.,
  {Merrifield} M.~R., {Kuijken} K., {Arnaboldi} M., {Gerhard} O., {Freeman}
  K.~C., 2005, \mnras, 357, 691

\bibitem[{{Navarro} {et~al.}(1995){Navarro}, {Frenk}, \& {White}}]{NFW95}
{Navarro} J.~F., {Frenk} C.~S., {White} S. D.~M., 1995, \mnras, 275, 720

\bibitem[{{Navarro} {et~al.}(1996){Navarro}, {Frenk}, \& {White}}]{NFW96}
---, 1996, \apj, 462, 563

\bibitem[{{Navarro} {et~al.}(1997){Navarro}, {Frenk}, \& {White}}]{NFW97}
---, 1997, \apj, 490, 493

\bibitem[{{Navarro} {et~al.}(2004){Navarro}, {Hayashi}, {Power}, {Jenkins},
  {Frenk}, {White}, {Springel}, {Stadel}, \& {Quinn}}]{Navarro+04}
{Navarro} J.~F., {Hayashi} E., {Power} C., {Jenkins} A.~R., {Frenk} C.~S.,
  {White} S.~D.~M., {Springel} V., {Stadel} J., {Quinn} T.~R., 2004, \mnras,
  349, 1039

\bibitem[{{O'Meara} {et~al.}(2001){O'Meara}, {Tytler}, {Kirkman}, {Suzuki},
  {Prochaska}, {Lubin}, \& {Wolfe}}]{O'Meara+01}
{O'Meara} J.~M., {Tytler} D., {Kirkman} D., {Suzuki} N., {Prochaska} J.~X.,
  {Lubin} D., {Wolfe} A.~M., 2001, \apj, 552, 718

\bibitem[{{Osipkov}(1979)}]{Osipkov79}
{Osipkov} L.~P., 1979, Soviet Astronomy Letters, 5, 42

\bibitem[{{O'Sullivan} {et~al.}(2003){O'Sullivan}, {Ponman}, \&
  {Collins}}]{OPC03}
{O'Sullivan} E., {Ponman} T.~J., {Collins} R.~S., 2003, \mnras, 340, 1375

\bibitem[{{Persic} {et~al.}(1996){Persic}, {Salucci}, \& {Stel}}]{PSS96}
{Persic} M., {Salucci} P., {Stel} F., 1996, \mnras, 281, 27

\bibitem[{{Pointecouteau} {et~al.}(2005){Pointecouteau}, {Arnaud}, \&
  {Pratt}}]{PAP05}
{Pointecouteau} E., {Arnaud} M., {Pratt} G.~W., 2005, \aap, 435, 1

\bibitem[{{Power} {et~al.}(2003){Power}, {Navarro}, {Jenkins}, {Frenk},
  {White}, {Springel}, {Stadel}, \& {Quinn}}]{Power+03}
{Power} C., {Navarro} J.~F., {Jenkins} A., {Frenk} C.~S., {White} S.~D.~M.,
  {Springel} V., {Stadel} J., {Quinn} T., 2003, \mnras, 338, 14

\bibitem[{{Prada} {et~al.}(2003){Prada}, {Vitvitska}, {Klypin}, {Holtzman},
  {Schlegel}, {Grebel}, {Rix}, {Brinkmann}, {McKay}, \& {Csabai}}]{Prada+03}
{Prada} F., {Vitvitska} M., {Klypin} A., {Holtzman} J.~A., {Schlegel} D.~J.,
  {Grebel} E.~K., {Rix} H.-W., {Brinkmann} J., {McKay} T.~A., {Csabai} I.,
  2003, \apj, 598, 260

\bibitem[{{Prugniel} \& {Simien}(1997)}]{PS97}
{Prugniel} P., {Simien} F., 1997, \aap, 321, 111

\bibitem[{{Rasia} {et~al.}(2004){Rasia}, {Tormen}, \& {Moscardini}}]{RTM04}
{Rasia} E., {Tormen} G., {Moscardini} L., 2004, \mnras, 351, 237

\bibitem[{{Rix} \& {White}(1992)}]{RW92}
{Rix} H., {White} S.~D.~M., 1992, \mnras, 254, 389

\bibitem[{{Romanowsky} {et~al.}(2003){Romanowsky}, {Douglas}, {Arnaboldi},
  {Kuijken}, {Merrifield}, {Napolitano}, {Capaccioli}, \&
  {Freeman}}]{Romanowsky+03}
{Romanowsky} A.~J., {Douglas} N.~G., {Arnaboldi} M., {Kuijken} K., {Merrifield}
  M.~R., {Napolitano} N.~R., {Capaccioli} M., {Freeman} K.~C., 2003, Science,
  301, 1696

\bibitem[{{S{\' a}iz} {et~al.}(2004){S{\' a}iz}, {Dom{\'{\i}}nguez-Tenreiro},
  \& {Serna}}]{SDS04}
{S{\' a}iz} A., {Dom{\'{\i}}nguez-Tenreiro} R., {Serna} A., 2004, \apjl, 601,
  L131

\bibitem[{{Saglia} {et~al.}(2000){Saglia}, {Kronawitter}, {Gerhard}, \&
  {Bender}}]{SKGB00}
{Saglia} R.~P., {Kronawitter} A., {Gerhard} O., {Bender} R., 2000, \aj, 119,
  153

\bibitem[{{Salucci} \& {Burkert}(2000)}]{SB00}
{Salucci} P., {Burkert} A., 2000, \apjl, 537, L9

\bibitem[{{Sanchis} {et~al.}(2004){Sanchis}, {{\L}okas}, \& {Mamon}}]{SLM04}
{Sanchis} T., {{\L}okas} E.~L., {Mamon} G.~A., 2004, \mnras, 347, 1198

\bibitem[{{Schwarzschild}(1979)}]{Schwarzschild79}
{Schwarzschild} M., 1979, \apj, 232, 236

\bibitem[{{Sersic}(1968)}]{Sersic68}
{Sersic} J.~L., 1968, Atlas de galaxias australes. Cordoba, Argentina:
  Observatorio Astronomico, 1968

\bibitem[{{Spergel} {et~al.}(2003){Spergel}, {Verde}, {Peiris}, {Komatsu},
  {Nolta}, {Bennett}, {Halpern}, {Hinshaw}, {Jarosik}, {Kogut}, {Limon},
  {Meyer}, {Page}, {Tucker}, {Weiland}, {Wollack}, \& {Wright}}]{Spergel+03}
{Spergel} D.~N., {Verde} L., {Peiris} H.~V., {Komatsu} E., {Nolta} M.~R.,
  {Bennett} C.~L., {Halpern} M., {Hinshaw} G., {Jarosik} N., {Kogut} A.,
  {Limon} M., {Meyer} S.~S., {Page} L., {Tucker} G.~S., {Weiland} J.~L.,
  {Wollack} E., {Wright} E.~L., 2003, \apjs, 148, 175

\bibitem[{{Stoehr}(2006)}]{Stoehr06}
{Stoehr} F., 2006, \mnras, 365, 147

\bibitem[{{Stoehr} {et~al.}(2002){Stoehr}, {White}, {Tormen}, \&
  {Springel}}]{SWTS02}
{Stoehr} F., {White} S.~D.~M., {Tormen} G., {Springel} V., 2002, \mnras, 335,
  L84

\bibitem[{{Treu} \& {Koopmans}(2002)}]{TK02}
{Treu} T., {Koopmans} L.~V.~E., 2002, \apj, 575, 87

\bibitem[{{Treu} \& {Koopmans}(2004)}]{TK04}
---, 2004, \apj, 611, 739

\bibitem[{{van der Marel}(1994)}]{vanderMarel94}
{van der Marel} R.~P., 1994, \mnras, 270, 271

\bibitem[{{Vikhlinin} {et~al.}(2005){Vikhlinin}, {Markevitch}, {Murray},
  {Jones}, {Forman}, \& {Van Speybroeck}}]{Vikhlinin+05}
{Vikhlinin} A., {Markevitch} M., {Murray} S.~S., {Jones} C., {Forman} W., {Van
  Speybroeck} L., 2005, \apj, 628, 655

\bibitem[{{Wilson} {et~al.}(2001){Wilson}, {Kaiser}, {Luppino}, \&
  {Cowie}}]{WKLC01}
{Wilson} G., {Kaiser} N., {Luppino} G.~A., {Cowie} L.~L., 2001, \apj, 555, 572

\bibitem[{{Wojtak} {et~al.}(2005){Wojtak}, {{\L}okas}, {Gottl{\"o}ber}, \&
  {Mamon}}]{WLGM05}
{Wojtak} R., {{\L}okas} E.~L., {Gottl{\"o}ber} S., {Mamon} G.~A., 2005, \mnras,
  361, L1

\bibitem[{{Yang} {et~al.}(2003){Yang}, {Mo}, \& {van den Bosch}}]{YMvdB03}
{Yang} X., {Mo} H.~J., {van den Bosch} F.~C., 2003, \mnras, 339, 1057

\bibitem[{{Young} \& {Currie}(1994)}]{YC94}
{Young} C.~K., {Currie} M.~J., 1994, \mnras, 268, L11

\bibitem[{{Zaritsky} \& {White}(1994)}]{ZW94}
{Zaritsky} D., {White} S.~D.~M., 1994, \apj, 435, 599

\bibitem[{{Zelen} \& {Severo}(1965)}]{ZS65}
{Zelen} M., {Severo} N.~C., 1965, Handbook of Mathematical Functions,
  {Abramowitz} M., {Stegun} I.~A., eds., Dover, New York, pp. 925--995

\end{thebibliography}
\end{document}